\documentclass[fleqn,usenatbib]{mnras}

\usepackage{newtxtext,newtxmath}

\usepackage[T1]{fontenc}

\DeclareRobustCommand{\VAN}[3]{#2}
\let\VANthebibliography\thebibliography
\def\thebibliography{\DeclareRobustCommand{\VAN}[3]{##3}\VANthebibliography}

\usepackage{graphicx}	
\usepackage{amsmath}	

\usepackage{orcidlink}

\newcommand{\aref}[1]{\hyperref[#1]{Appendix~\ref{#1}}}

\title[CONGRuENTS I]{\textsc{CONGRuENTS} (COsmic-ray, Neutrino, Gamma-ray and Radio Non-Thermal Spectra). I. A predictive model for galactic non-thermal emission}

\author[M. A. Roth et al.]{
Matt A. Roth,$^{\orcidlink{0000-0002-4204-5026}}$$^{1}$\thanks{E-mail: matt.roth@anu.edu.au (MAR)}
Mark R. Krumholz$^{\orcidlink{0000-0003-3893-854X}}$$^{1,2}$,
Roland M. Crocker$^{\orcidlink{0000-0002-2036-2426}}$$^{1}$,
and Todd A. Thompson$^{\orcidlink{0000-0003-2377-9574}}$$^{3}$
\\
$^{1}$Research School of Astronomy \& Astrophysics, The Australian National University, Canberra, Australian Capital Territory, 2611, Australia\\
$^{2}$ARC Centre of Excellence for All-Sky Astrophysics in
Three Dimensions (ASTRO-3D), Canberra, Australian Capital Territory, 2611, Australia\\
$^{3}$Department of Astronomy and Center for Cosmology \& Astro-Particle Physics, The Ohio State University, Columbus, OH 43210, USA
}

\date{Accepted XXX. Received YYY; in original form ZZZ}

\pubyear{2022}

\begin{document}
\label{firstpage}
\pagerange{\pageref{firstpage}--\pageref{lastpage}}
\maketitle

\begin{abstract}
The total luminosity and spectral shape of the non-thermal emission produced by cosmic rays depends on their interstellar environment, a dependence that gives rise to correlations between galaxies' bulk properties -- star formation rate, stellar mass, and others -- and their non-thermal spectra. Understanding the physical mechanisms of cosmic ray transport, loss, and emission is key to understanding these correlations. Here, in the first paper of the series, we present a new method to compute the non-thermal spectra of star-forming galaxies, and describe an open-source software package -- COsmic-ray, Neutrino, Gamma-ray and Radio Non-Thermal Spectra (\textsc{CONGRuENTS}) -- that implements it. 
As a crucial innovation, our method requires as input 
only a galaxy's effective radius, star formation rate, stellar mass, and redshift, all quantities that are readily available for large samples of galaxies and do not require expensive, spatially resolved gas measurements. From these inputs we derive individual, galaxy-by-galaxy models for the background gas and radiation field through which cosmic rays propagate, from which we compute steady state cosmic ray spectra for hadronic and leptonic particles in both the galactic disc and halo by solving the full kinetic equation. We invoke modern models for cosmic ray transport and include all significant emission and loss mechanisms. 
In this paper we describe the model and validate it against non-thermal emission measured in nearby star-forming galaxies that span four orders of magnitude in star formation rate.

\end{abstract}




\section{Introduction}

Star-formation is a prolific and often dominant source of cosmic rays (CRs), non-thermal particles believed to be predominantly accelerated in the shock waves of supernova remnants. This acceleration process yields a particle distribution in the form of a power law in particle momentum \citep{1978MNRAS.182..147B, 1987PhR...154....1B} $dN/dp \propto p^{-q}$, where $q$ is the injection index.
Subsequent to escape from their acceleration region, 
the generally energy-dependent processes of cooling and transport 
through the interstellar medium set in
to shape the CR distribution.
While at low energies ionisation losses significantly dominate the overall loss rate for CR electrons and positrons (hereafter, `electrons'), at higher energies inverse Compton and synchrotron losses become the important cooling channels. At intermediate energies, bremsstrahlung and diffusive losses also become competitive in some systems. For hadrons the picture is somewhat simpler, with only two relevant competing loss mechanisms in the form of hadronic collisions that form pions and diffusive escape; ionisation losses only become significant at proton energies below the pion production threshold, and therefore below the energy at which protons produce observable non-thermal emission.
The transition between these loss regimes is imprinted in the steady-state cosmic ray spectrum and affects the non-thermal spectra derived from such a population.

Accurate modelling of these cooling mechanisms is challenging because they depend not just on the energies of individual CRs, but also on properties of the galactic environment. Relevant properties include the interstellar and circumgalactic gas density and magnetic field strength, and the intensity and spectral shape of the interstellar radiation field (ISRF).

Given this complication, detailed models of the non-thermal spectra of galaxies have 
tended to adopt one of two strategies. The first is a `single zone' approach whereby one approximates the galaxy as a single, uniform medium \citep[e.g.,][]{Akyuz1991, Sreekumar1994,Volk1996,Paglione1996, Romero2003,Torres2004,Domingo-Santamaria2005,deCeadelPozo2009, Rephaeli2010}.
\citet{2010ApJ...717....1L} pioneered the use of this approach in the context of a systematic study
of non-thermal emission across the star-forming main sequence. While conceptually simple, a disadvantage of this method is that it typically relies on a large suite of observational inputs to characterise mean environmental parameters, for example gas density and interstellar radiation field strength, important to shaping the non-thermal radiation \citep[e.g.,][]{Yoast-Hull13a, Yoast-Hull16a, Wang18a, 2018PASJ...70...49S, 2019MNRAS.487..168P, 2020MNRAS.493.2817K}.

The second approach is detailed 3D models or numerical simulations that follow both the CRs and the gas through which they move \citep[e.g.,][]{Ackermann12a, Johannesson16a, Werhahn21a, Werhahn21b, Werhahn21c, Hopkins22a}.
In principle this latter approach should provide a higher  level of accuracy than the single zone technique.
For sufficiently nearby systems, moreover, 
these detailed predictions can be then be compared against observation.
However, the detailed, 3D modelling approach is not easily scalable, at least for now, given that 3D simulations are computationally expensive, and that direct observations of the 3D structure of the gas density, magnetic field strength, and ISRF are extraordinarily difficult even in nearby galaxies, and completely inaccessible in large samples of galaxies beyond the local Universe. This practical consideration means that accurate models capable of predicting CR emission in terms of data that are more accessible -- for example, star formation rates and total stellar masses -- are key to understanding the population-level connections between non-thermal emission in different spectral bands and other galaxy properties, as embodied in empirical relations such as the FIR-radio correlation \citep{1989A&A...218...67V, 1990MNRAS.245..101C, Condon92a, Bell03a, Brown17a}, the FIR-$\gamma$ correlation \citep{Ajello20a, 2020A&A...641A.147K}, While there have been fruitful theoretical attempts to address these large-scale correlations \citep[e.g.,][]{Thompson06a, 2010ApJ...717....1L, Lacki10b, Schober16a}, these generally adopted  simple models for the background galaxy state that have a limited range of applicability. Here, extending previous single zone approaches, we present a new method to predict the broad-band, non-thermal spectra of individual star-forming galaxies and calculate CR and non-thermal spectra self-consistently. 
Our model takes careful account of recent advances in the understanding of cosmic ray transport and makes use only of galactic star formation rates, stellar masses, sizes, and redshifts. These quantities are all readily accessible observationally, even at moderate to high redshift.
There are, moreover, 
samples of tens of thousands of individual objects for which these quantities are readily available \citep[e.g.,][]{van-der-Wel12a, van-der-Wel14a}. 
We use these quantities as inputs to a detailed and accurate treatment of all microphysical loss processes.
This allows us to predict the full CR proton and electron distributions --
both within a galactic disc and in its circumgalactic halo
-- and the emission they produce. 

A preliminary version of our computational method was verified
in \citep{2021Natur.597..341R} wherein we demonstrated that star-forming galaxies seem to explain the isotropic, diffuse $\gamma$-ray flux observed from the universe in the {\it Fermi}-LAT band.
In this paper, we verify our extended method by direct comparison to a number of nearby star-forming galaxies that have been observed from the radio to $\gamma$-rays and whose spectra 
do not appear to be significantly polluted by emission from active galactic nuclei. We show that our model performs well over a very wide range of absolute and specific star-formation rates (sSFR), from Arp 220 ($\mathrm{SFR}\approx 220$ M$_{\odot}$ yr$^{-1}$, $\log{\mathrm{sSFR}/\mbox{yr}^{-1}} \approx -8.29$) down to M31 ($\mathrm{SFR}\approx 0.26$ M$_{\odot}$ yr$^{-1}$, $\log{\mathrm{sSFR}/\mathrm{yr}^{-1}} \approx -11.4$). In the second paper in this series (Roth et al., in prep.), we apply our model to large galaxy catalogues in order to explain the origin of the various population-level correlations mentioned above.

Our plan for the remainder of this paper is as follows. We first discuss in \autoref{sec:galaxy_properties} how we obtain the galaxy properties required to calculate CR losses from easily observable data.
We follow in \autoref{sec:cr_spectra} with a description of our method for computing the steady-state spectra for CR protons and electrons. In \autoref{sec:non-thermal} we explain how we compute non-thermal emission rates from the steady-state spectra. We compare the results of our models to observations in \autoref{sec:comparison}, and summarise and discuss future plans and prospects in \autoref{sec:conclusion}.

\section{Galaxy properties}
\label{sec:galaxy_properties}

We extend the model of \citet{2021Natur.597..341R} to obtain a basic two zone model for a galactic disc and halo. We take as input the stellar mass $M_{\ast}$, the star formation rate $\dot{M}_{\ast}$, and the effective or half-light radius $R_{\rm e}$, supplemented by a redshift $z$. We purposefully limit ourselves to this basic set of parameters as they can be measured for a large number of galaxies out to high redshift. We model galaxies as plane-parallel slabs consisting of a disc (\autoref{ssec:disc}) and a halo (\autoref{ssec:halo}), both pervaded by an interstellar radiation field (ISRF; \autoref{ssec:isrf}).

\subsection{The disc}
\label{ssec:disc}

The first component in our model is a galactic disc, which we characterise by midplane values of the number density of H nuclei $n_{\rm H}$, ionisation fraction by mass $\chi$, gas velocity dispersion $\sigma_{\rm g}$, gas scale height $h_{\rm g}$, and magnetic field amplitude $B$; there is also an ISRF, discussion of which we defer to \autoref{ssec:isrf}. We derive these quantities from our basic observables -- $M_{\ast}$, $\dot{M}_{\ast}$, and $R_{\rm e}$ -- using a mix of empirical correlations and simple physical arguments.

The first step in our procedure is to derive the gas surface density and velocity dispersion from two empirical correlations. For the first of these we invert the Extended Schmidt Law obtained by \citet{2011ApJ...733...87S}. 
\begin{equation}
\Sigma_{\rm g} = 10^{4.28} \left(\frac{\dot{\Sigma}_{\ast}}{\rm M_{\odot}\, Myr^{-1} \,pc^{-2}} \right) \left( \frac{\Sigma_{\ast}}{\rm M_{\odot} \, pc^{-2}} \right)^{-0.48} \, \frac{\rm M_{\odot}}{\rm pc^{2}},
\end{equation}
where $\Sigma_* = M_* / 2\pi R_{\rm e}^2$ and $\dot{\Sigma}_* = \dot{M}_* / 2\pi R_{\rm e}^2$ are the stellar mass and star formation rate per unit area, respectively. For the second, we use a linear fit to the correlation shown in Figure 7 of \citet{2019MNRAS.486.4463Y}, which yields
\begin{equation}
    \sigma_{\rm g} = 39.81 \left( \frac{\dot{M}_{\ast} }{\mathrm{M}_\odot\,\mathrm{yr}^{-1}} \right)^{0.2} \, {\rm km \, s^{-1}}.
\end{equation}

We next compute the gas scale height by assuming that the gas is in hydrostatic equilibrium. This requires some care, because the contribution of stars to the gravitational potential felt by the gas depends on the ratio of gas and stellar scale heights. We therefore adopt the approximation proposed by \citet{Forbes12a},
\begin{equation}
    h_{\rm g}\approx \frac{\sigma_{\rm g}^2}{\pi G \left[\Sigma_{\rm g} + \left( \sigma_{\rm g} / \sigma_{\ast} \right) \Sigma_{\ast} \right]}
\end{equation}
where $G$ is the gravitational constant and $\sigma_{\ast}$ is the stellar velocity dispersion.
If the gas and stars are well-mixed, as we expect to be approximately the case in starburst and high-redshift galaxies where the gas velocity dispersion is large, 
then we have $\sigma_{\rm g} / \sigma_{\ast} \sim 1$, but if the stellar scale height is much larger than the gas scale height, as is the case for most gas-poor, low velocity dispersion modern galaxies, $\sigma_{\ast}$ is expected to be much larger than $\sigma_{\rm g}$ and hence the contribution of $\Sigma_{\ast}$ to maintaining the equilibrium is likewise much smaller,
since any parcel of gas is confined only by the gravity of stars closer to the midplane than it  (i.e. $\sigma_{\rm g} / \sigma_{\ast} \ll 1$).
We take our values of $\sigma_{\ast}$ from the empirical correlation obtained by \citet{2012ApJ...760...62B}, who derive an inferred (stellar) velocity dispersion of the form $\sigma_{\ast} = \sqrt{G M_{\ast} / \left[ 0.557 K_{\nu}\!\left(n\right) R_{\rm e} \right]}$, where $K_{\nu}\!\left(n\right) = 73.32/[10.465 + \left( n - 0.94 \right)^{2}] + 0.954$ is a virial constant \citep{2002A&A...386..149B} that depends on the S\'ersic index $n$. We set $n=1$ as appropriate for spiral galaxies.

Once we have determined the scale height, we compute the ISM hydrogen number density as
\begin{equation}
    n_{\rm H} = \frac{\Sigma_{\rm g}}{2 \mu_{\rm H} m_{\rm H} h_{\rm g}}
\end{equation}
where $\mu_{\rm H} = 1.4$ is the mean mass per hydrogen nucleus in units of the hydrogen mass $m_{\rm H}$ for the usual cosmic mix of 74\% H, 25\% He by mass. We assume that the ionisation fraction by mass in this gas is $\chi = 10^{-4}$; this value is expected in rapidly star-forming galaxies based on astrochemical modelling \citep{2020MNRAS.493.2817K}, and is also intermediate between the ionisation fractions $\approx 10^{-2}$ and $\approx 10^{-6}$ found in the atomic and molecular phases of more slowly star-forming galaxies like the Milky Way. We do not attempt a more accurate estimate of $\chi$ because \citet{2021Natur.597..341R} show that its effects are minimal.

Finally, we also estimate the magnetic field strength following \citet{2021Natur.597..341R}, who note that dynamo processes tend to force magnetic field strengths in galaxies to lie in a fairly narrow range of Alfv\'en Mach number $\mathcal{M}_A$ (also see the discussion in \citealt{2022arXiv220213020B}). For \textsc{CONGRuENTS} we adopt a fiducial value $\mathcal{M}_A = 2$. Assuming equipartition between turbulent modes on large scales (see simulations by e.g. \citealt{2016JPlPh..82f5301F}), the velocity dispersion in Alfv\'en modes is $\mathcal{M}_A v_A = \sigma_{\rm g}/\sqrt{2}$  as adopted by \citet{2020MNRAS.493.2817K}, where $v_A = B/\sqrt{4\pi n_{\rm H} \mu_{\rm H} m_{\rm H}}$ is the Alfv\'en speed, and therefore
\begin{equation}
    B = 3.84 \,\mathcal{M}_A^{-1} \left(\frac{\sigma_{\rm g}}{10\mbox{ km s}^{-1}}\right) \left(\frac{n_{\rm H}}{1\mbox{ cm}^{-3}}\right)^{1/2}\,\mu\mathrm{G}.
\end{equation}
This completes specification of all the midplane quantities in the disc.

\subsection{The halo}
\label{ssec:halo}

The second 
environment treated in
our model is a low-density halo that exists outside the disc; this component is required because, while losses from cosmic ray protons occur primarily in the disc (simply because the disc dominates the total gas mass and thus the set of available targets for hadronic interactions), electron losses can also be driven by magnetic and radiation fields present in regions of very low matter density. Indeed, observations of edge-on galaxies show that synchrotron emission has a substantially larger scale height than gas \citep{Krause18a}. On the other hand, the scale heights of synchrotron emission are still significantly smaller than galactic scale lengths, and thus it is reasonable to continue to use a plane-parallel geometry even in the halo.

We set the density of gas in the halo to $10^{-3}$ of the disc value, and the scale height of the halo to $50 h_g$. In practice, both these choices make almost no difference to the final result, as we now explain: 
Our choice of density is motivated by simulations by \citet{2021arXiv210504136W}. 
For any reasonable choice of density in the halo away from the inner disc, however, losses due to collisional interactions with matter (hadronic interactions, bremsstrahlung, and ionisation or Compton losses)
are only relevant at the very lowest energies where they have negligible impact on the non-thermal emission.
Similarly, we adopt a scale height because one is required in our numerical method (see below), but our choice is essentially a placeholder, as the scale height matters only for diffusive losses, which are always negligible for CR electrons in the halo. We therefore only require a value of $h_g$ large enough to guarantee this condition. 
We emphasise that these obviously oversimplified choices are justified only because we are interested only in the integrated non-thermal emission, not in actually computing the spatial distribution of emission; doing would require
a more detailed model that accounts for the decay of the matter density with height. However, as we will discuss now, we are essentially only interested in the balancing of synchrotron (approximately $\propto u_{\rm mag}$) and inverse Compton losses (approximately $\propto u_{\rm rad}$), which are the dominant loss processes that produce non-thermal emission.

We set the halo magnetic field amplitude to one third the disc value, $B/3$, based on simulations \citep{2021arXiv210504136W} for vigorously star-forming systems.
For more quiescent systems we set it to $B/1.5$  based on observational results that show the decay of the magnetic field amplitude to be less pronounced in weakly star-forming galaxies \citep{2013A&A...560A..42M, 2018A&A...615A..98M}; we place the cut between these two regimes at a specific star-formation rate of $\log (\mathrm{sSFR}/{\rm yr}^{-1}) = -10$. We also adopt a plane parallel approximation whereby the ISRF in the halo (\autoref{ssec:isrf}) is the same as that in the plane. Note that, under our assumption that collisional losses and diffusive escape are unimportant, these choices for the halo matter only to the extent that they specify what fraction of halo losses will be into synchrotron versus inverse Compton radiation, and thus the balance between halo radio and $\gamma$-ray emission.

\subsection{Interstellar radiation field}
\label{ssec:isrf}

Both the electron energy loss rate and the spectrum of photons emitted in the inverse Compton process depend on the spectrum of the ISRF that provides the target photons for the scattering interactions. While some authors approximate the spectrum as monochromatic \citep[e.g.,][]{2019MNRAS.487..168P} or single-component \citep[e.g.,][]{Yoast-Hull13a} for this purpose, this approach cannot be applied over a broad range of galaxy properties, and adopting it may hide important features in the derived inverse Compton spectrum. In particular, the ISRF energy density in dust-poor galaxies is dominated by a direct starlight component with an effective temperature of thousands of K, while in dust-rich starbursts it is dominated by an infrared component with an effective temperature of tens of K.
This shift in the peak frequency of the ISRF has important consequences for the frequencies at which inverse Compton emission from these galaxies emerges, and, thanks to the energy-dependence of the inverse Compton loss rate in the Klein-Nishina regime, the partition of electron losses between synchrotron and inverse Compton channels. We thus use a more realistic model for the ISRF that is individually determined for each galaxy.
Our model consists of six components: a black body spectrum for the cosmic microwave background at a temperature of $T_{\rm CMB} = (1+z) \; 2.725 \; {\rm K}$, a dilute modified black body for far-infrared dust emission, and three dilute black bodies at 3000 K, 4000 K and 7500 K plus a broken power law model for the FUV radiation  
with the shape given by \citet[see also \citealt{2011piim.book.....D}]{Mathis83a} to represent the starlight field. The photon number density distribution for the black body components has the usual functional form
\begin{equation}
    \frac{dn_{\rm ph}}{dE_{\rm ph}} \bigg|_{{\rm BB},T} =
    \frac{8 \pi E_{\rm ph}^2}{\left( h c \right)^3} \frac{1}{\exp{\left(E_{\rm ph}/k_{\rm B} T\right)}-1},
\end{equation}
where $T$ is the black body temperature, $h$ the Planck constant, $k_{\rm B}$ the Boltzmann constant, and $c$ the speed of light, while for the modified black body we use the functional form proposed by \citet{2002ApJ...568...88Y} and \citet{2008A&A...486..143P},
\begin{equation}
\frac{dn_{\rm ph}}{dE_{\rm ph}} \bigg|_{{\rm mod BB},T} = \left(\frac{E_{\rm ph}}{E_0}\right) \frac{dn_{\rm ph}}{dE_{\rm ph}} \bigg|_{{\rm BB},T} ,
\end{equation}
where $E_0/h = 2000$ GHz. We set the temperature $T$ of the modified black body using the empirical scaling relation for dust effective temperature obtained by \citet{2014A&A...561A..86M},
\begin{equation}
T_{\rm dust} = \left[ 98 \left( 1+z \right)^{-0.065} + 6.9 \log_{10}\left( \frac{\dot{M_{\ast}}}{M_{\ast}} {\rm yr}\right) \right] \; {\rm K} \, .
\end{equation}
Thus our total expression for the ISRF photon number density takes the form
\begin{eqnarray}
    \frac{dn_{\rm ph}}{dE_{\rm ph}} & = & \frac{dn_{\rm ph}}{dE_{\rm ph}} \bigg|_{{\rm BB},T_{\rm CMB}} + C_{\rm dil, dust} \frac{dn_{\rm ph}}{dE_{\rm ph}} \bigg|_{{\rm modBB},T_{\rm dust}} + {}
    \nonumber \\
    & & 
    C_{\rm dil, 3K} \frac{dn_{\rm ph}}{dE_{\rm ph}} \bigg|_{{\rm BB},3000\,\mathrm{K}}+ C_{\rm dil, 4K} \frac{dn_{\rm ph}}{dE_{\rm ph}} \bigg|_{{\rm BB},4000\,\mathrm{K}} + {}
    \nonumber \\
    & &
    C_{\rm dil, 7.5K} \frac{dn_{\rm ph}}{dE_{\rm ph}} \bigg|_{{\rm BB},7500\,\mathrm{K}} + C_{\rm dil, FUV} \frac{dn_{\rm ph}}{dE_{\rm ph}} \bigg|_{{\rm FUV}},
    \label{eq:dndEp}
\end{eqnarray}
where the factors $C_{\rm dil}$ are the dilution factors of the various components, and $dn_{\rm ph}/dE_{\rm ph}|_{\rm FUV}$ is the photon number density distribution corresponding to the energy distribution given by equation 12.7 of \citet{2011piim.book.....D}.
To obtain the dilution factors, we associate each term in \autoref{eq:dndEp} with a particular source of radiant luminosity $L$ for the galaxy. If we assume that this radiates from the mid-plane over a region of effective radius $R_{\rm e}$, we then have $C_{\rm dil} = {L}/{u_{\rm rad} 2 \pi R_{\rm e}^2 c}$. Here $u_{\rm rad}$ is the radiation energy density corresponding to the undiluted field, which is $u_{\rm rad, BB} = a_{\rm rad} T^4$ for the black body, where $a_{\rm rad}$ is the radiation constant, and
\begin{equation}
u_{\rm rad, modBB} = 24.89 \frac{8 \pi}{\left( h c \right)^3} \frac{\left(  k_{\rm B} T\right)^5}{E_0}
\end{equation}
for the modified black body. Thus our final step is to associate a luminosity to each component, and express that luminosity in terms of our input quantities, $M_\ast$ and $\dot{M}_\ast$.

To obtain the luminosities for each spectral component we follow the approach of DustPedia \citep{2019A&A...624A..80N} by dividing galactic luminosities into four components: light from old (age $\gtrsim 200$ Myr) and young stars, each of which contains a part that is absorbed by dust and a part that is unattenuated and thus can be observed directly; we refer to these components as $L_{\rm abs,young}$, $L_{\rm abs,old}$, $L_{\rm obs,young}$, and $L_{\rm obs,old}$, respectively.

\begin{figure}
	\includegraphics[width=\columnwidth]{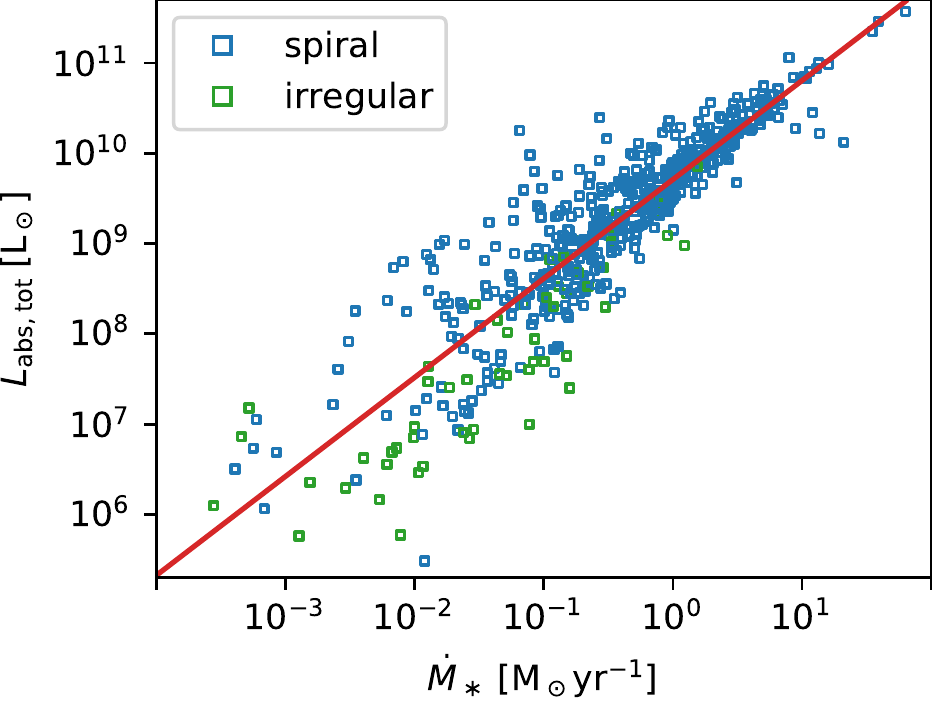}
    \caption{Total dust-absorbed luminosities estimated for spiral and irregular galaxies in DustPedia \citep[blue and green points;][]{2019A&A...624A..80N}, together with our power law fit (red line; \autoref{eq:Labstot}).}
    \label{fig:Labstot}
\end{figure}

\begin{figure}
	\includegraphics[width=\columnwidth]{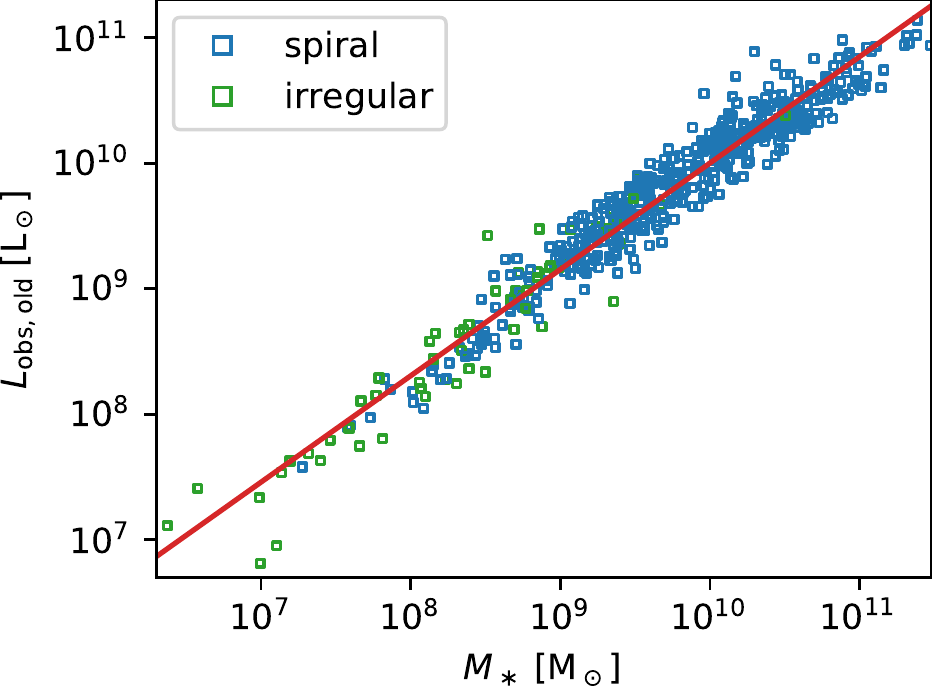}
    \caption{Observed luminosity of the old stellar population in spiral and irregular galaxies in DustPedia \citep[blue and green points;][]{2019A&A...624A..80N}, together with our power law fit (red line; \autoref{eq:Lobsold}).}
    \label{fig:Lobsold}
\end{figure}

\begin{figure}
	\includegraphics[width=\columnwidth]{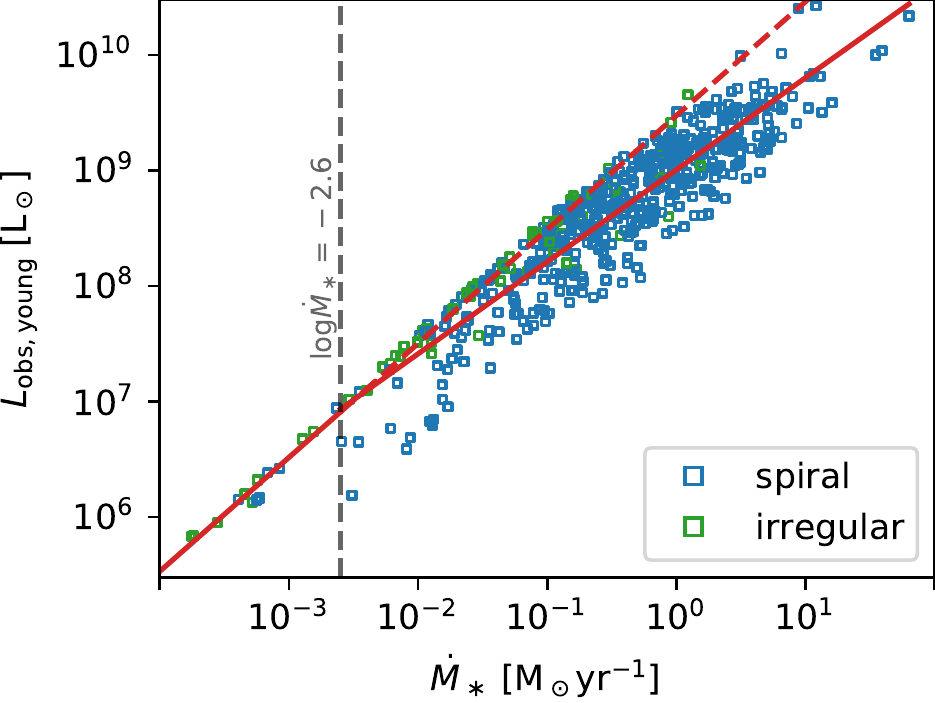}
    \caption{Observed luminosity of the young stellar population in spiral and irregular galaxies in DustPedia \citep[blue and green points;][]{2019A&A...624A..80N}, together with our broken power law fit (solid red line; \autoref{eq:Lobsyoung}). The break accounts for the saturation of $L_{\rm obs,young}$ to a maximum possible value in quiescent, dust-poor systems where none of the starlight is absorbed by dust; the dashed red line illustrates this upper envelope, and is obtained simply by extending our fit for $\log(\dot{M}_\ast/{\rm M}_\odot\,{\rm yr}^{-1})<-2.6$ to higher SFR.
    }
    \label{fig:Lobsyoung}
\end{figure}

We assume that the 3000 and 4000 K dilute black bodies are powered by the unattenuated old stellar population $L_{\rm obs, old}$, with $L_{3000\,\mathrm{K}} = 0.57 L_{\rm obs,old}$ and $L_{4000\,\mathrm{K}} = 0.43 L_{\rm obs,old}$; the factors of $0.57$ and $0.43$ are set to be equal to the ratio of these two components in the fit provided by \citet{2011piim.book.....D} for the Milky Way's ISRF. We assume that these factors are the same for all systems because they primarily depend on the initial mass function, which sets the ratio of K stars (effective temperature $\sim 4000$ K) to M stars (effective temperature $\sim 3000$ K). A more detailed model would also include the effects of the star-formation history (SFH), but these are not major, since red colours evolve only very slowly with time for older stellar populations. Similarly, we assume that the 7500 K dilute black body and the FUV field are supplied by the unattenuated young stars, $L_{\rm obs, young}$, with $L_{7500\,\mathrm{K}} = 0.76 L_{\rm obs,young}$ and $L_{\rm FUV} = 0.24 L_{\rm obs,young}$; again, the numerical factors are the same as those provided by \citeauthor{2011piim.book.....D} for the Milky Way. These factors are somewhat more uncertain, since they likely do depend on both the star formation rate and the amount of dust extinction in the galaxy; however, in galaxies with high star formation rates and a great deal of dust extinction, the 7500 K and FUV components tend to be highly sub-dominant in any event, so we do not attempt to model this effect more carefully. Finally, the dust emission consists of the portion of the (old and young) luminosity that is absorbed and subsequently re-emitted in the FIR by dust grains, i.e. $L_{\rm dust} = L_{\rm abs,tot} = L_{\rm abs,old} + L_{\rm abs, young}$. 
To obtain the various luminosities in terms of $M_\ast$ and $\dot{M}_\ast$, we perform simple power law fits to the DustPedia data set \citep{2019A&A...624A..80N}, which provides estimates of each luminosity component derived by fitting the observed optical plus IR spectral energy distributions using the CIGALE code \citep{Boquien19a}; we exclude  
elliptical and lenticular (S0) galaxies from our sample, leaving spiral and irregular galaxies to obtain our relations. For the FIR emission (which we fit using the sum of the old and young absorbed components), we find a best fit
\begin{equation}
    \frac{L_{\rm abs, tot}}{10^9 \,\rm L_\odot} = 5.13 \left( \frac{\epsilon \dot{M_{\ast}}}{\rm M_\odot \,yr^{-1}} \right)^{1.10},
    \label{eq:Labstot}
\end{equation}
where we have introduced $\epsilon = 1.79$ to convert from a Chabrier to a Salpeter IMF \citep{2007MNRAS.378.1550P}.We show this fit together with the data in \autoref{fig:Labstot}. For the unattenuated old stellar population, our fit is 
\begin{equation}
    \frac{L_{\rm obs, old}}{10^9\,\rm L_\odot} = 1.48 \left( \frac{\epsilon M_{\ast}}{10^9\, \rm M_\odot} \right)^{0.85},
    \label{eq:Lobsold}
\end{equation}
and we show the fit and data together in \autoref{fig:Lobsold}. Finally, for the young component, we observe that the data are better fit by a two-component power law than a single one, with the low-luminosity portion of the power law likely representing galaxies with so little dust content that absorption is negligible. Our fit for this quantity is
\begin{equation}
    \frac{L_{\rm obs, young}}{10^9\,\rm L_\odot} =
    \begin{cases} 
    1.02 \left( \frac{\epsilon \dot{M_{\ast}}}{\rm M_\odot yr^{-1}} \right)^{0.797}, & {\rm if } \log\left(\frac{\epsilon \dot{M_{\ast}}}{{\rm M}_\odot\,{\rm yr}^{-1}}\right) > -2.6\\
    2.30 \left( \frac{\epsilon \dot{M_{\ast}}}{\rm M_\odot yr^{-1}} \right)^{0.987}, & {\rm otherwise}.
    \end{cases}
    \label{eq:Lobsyoung}
\end{equation}
We show the fit and data in \autoref{fig:Lobsyoung}.

\section{CR spectra}
\label{sec:cr_spectra}

Now that we have specified our model for the background galaxy, the next step in \textsc{CONGRuENTS} is to calculate steady-state CR spectra by balancing CR injection and loss. We describe our treatment of CR injection in \autoref{ssec:cr_injection}, our method for calculating proton spectra in \autoref{ssec:proton_spectra}, and our method for electron spectra in \autoref{ssec:electron_spectra}.

\subsection{CR injection}
\label{ssec:cr_injection}

Our treatment of CR injection follows the approach described in \citet{2021Natur.597..341R}. We refer readers to that paper for full details, but to summarise here, we assume that a fraction of the kinetic energy injected into the interstellar medium by supernovae is ultimately deposited in non-thermal particles in a process of diffusive shock acceleration, yielding a particle distribution in the form of a power law in particle momentum $d\dot{N}_{\rm CR}/dp \propto p^{-q}$, where $q$ is the injection index. Test particle analytical calculations for this process yield a spectrum with $q = 2.0$, whereas observational results suggest a CR injection index between $\approx 2.1$ to $\approx 2.5$.
We adopt $q=2.2$ as our fiducial choice  \citep{2011JCAP...05..026C,2012JCAP...07..038C}. We normalise the injection rate by assuming that 10\% of supernova kinetic energy is deposited in CR protons, equivalent to $10^{50}$ erg per SN \citep{1995STIN...9622970W,2013A&A...553A..34D}, ignoring the smaller fraction of heavier ions, and a further 2\% ($1 \times 10^{49}$ erg) in primary cosmic ray electrons \citep{2010ApJ...717....1L}. We note here that our choice of normalisation, while a standard assumption and shown to be successful -- see for example \citet{2010ApJ...717....1L} and \citet{2021Natur.597..341R} -- has some uncertainty attached to it. For instance \citet{2020MNRAS.493.2817K} required a proton injection fraction of 5\% for M82. Under these assumptions, \citet{2021Natur.597..341R} shows that the particle injection rate per unit energy is \begin{equation}
    \frac{d\dot{N}_{\rm CR}}{dE_{\rm CR}} = \Phi \dot{M}_{\ast} \left(\frac{p}{p_{\rm 0}}\right)^{-q} \frac{dp}{dE_{\rm CR}} \; {\rm e}^{-E_{\rm CR}/E_{\rm cut}}
    \label{eq:cr_inject}
\end{equation}
For cosmic ray protons, the normalisation $\Phi = 6.22 \times 10^{42} \, {\rm s^{-1} \, GeV^{-1} \, M_\odot^{-1} \, yr}$ with $p_{\rm 0} = 1\;{\rm GeV}/c$, and we adopt a cutoff energy $E_{\rm cut} = 100 \, {\rm PeV}$; only the ultra-high energy neutrino spectrum is sensitive to $E_{\rm cut}$, due to the effects of $\gamma \gamma$ opacity on $\gamma$-rays of similar energy \citep{2021Natur.597..341R}. For cosmic ray electrons, we have $\Phi = 2.98 \times 10^{34} \, {\rm s^{-1} \, GeV^{-1} \, M_\odot^{-1} \, yr}$ with $p_{\rm 0} = 1\;{\rm MeV}/c$, and a lower cutoff energy $E_{\rm cut} = 100 \, {\rm TeV}$.
%
%

\subsection{Cosmic ray protons}
\label{ssec:proton_spectra}

Loss processes for CR electrons are varied and strongly dependent on the galactic environment.
On the other hand,
as discussed in more detail in the following sections, CR protons  predominantly suffer two fates: a) they survive long enough to diffusively escape into the halo, where loss times become of order the Hubble time or longer, or b) they suffer hadronic collisions and lose energy by producing mesons. Other loss mechanisms are insignificant for CRs with energies above the pion production threshold, which are the only protons of interest here, since only they produce observable non-thermal emission. To determine the fraction of cosmic ray protons that lose their energy in these hadronic collisions rather than escaping, we adopt as a fiducial choice the model for CR transport proposed by \citet{2020MNRAS.493.2817K}. We choose this model because it is physically-motivated rather than purely phenomenological, and has been shown to provide good predictions for the $\gamma$-ray spectral shapes of nearby starburst galaxies \citep{2020MNRAS.493.2817K} and of the integrated cosmological gamma-ray background \citep{2021Natur.597..341R}. However, we caution that no CR transport model to date successfully reproduces all 
the CR phenomenology of the Milky Way -- see \citet{2022MNRAS.514..657K} and \citet{2022MNRAS.517.5413H} for more discussion. For this reason, the \textsc{CONGRuENTS} code allows users the option of varying this model and adopting a modified prediction for CR transport, as we discuss in more detail below.

The \citet{2020MNRAS.493.2817K} model suggests that within the predominantly-neutral disc of the galaxy, CR transport is governed by the balance of the growth rate of the streaming instability excited by cosmic rays and ion-neutral damping. This leads to diffusion by streaming along magnetic field lines and field line random walk processes, which can be approximated, at sufficiently large scales, by an energy dependent diffusion coefficient in the form of
\begin{equation}
    D\left( E_{\rm p}\right) \approx V_{\rm st} \frac{h_{\rm g}}{\mathcal{M}_{\rm A}^3},
    \label{eq:diffcoef}
\end{equation}
where the streaming velocity $V_{\rm st}$ is given by\footnote{The halo gas is fully ionised, hence damping by ion neutral damping is not applicable. However, the value of the diffusion coefficient in the halo is unimportant. This is a significant contrast between our model and those adopted by some earlier authors, e.g., \citet{2001ApJ...547..264J}. In these earlier works, CR transport in the halo is assumed to be described purely by isotropic diffusion, in which case CRs in the halo can random walk back into the disc; if this possibility exists, the diffusion coefficient in the halo does matter. By contrast, we assume that CR transport is primarily by streaming, for which diffusion is only an approximate description; in this case CRs that reach the halo do not re-enter the disc, because the strong CR pressure gradient outward from the disc causes them to stream away. Since we further assume that diffusive escape of leptons from the halo is negligible, and diffusive escape of protons does not matter because halo protons do not produce observable emission, the exact value of the diffusion coefficient in the halo does not matter at all for us.}
\begin{equation}
    V_{\rm st} \approx \min\left[c,
V_{\rm Ai} \left(
1 + \frac{\gamma_d \; \chi \; \mathcal{M}_{\rm A} \; c \; \mu_{\mathrm{H}}^{3/2} m_\mathrm{H}^{3/2} n_\mathrm{H}^{3/2}}{4 \; \pi^{1/2} \; C \; e \; u_{\rm LA} \; \mu_{\rm i} \; \gamma^{-q+1}}
\right)
\right].
\end{equation}
Here, $V_{\rm Ai}$ is the ion Alfv\'en speed given by $V_{\rm Ai} = u_{\rm LA}/\left(\chi^{1/2} \mathcal{M}_{\rm A}\right)$, $u_{\rm LA} = \sigma_\mathrm{g}/\sqrt{2}$ is the velocity dispersion in Alfv\'en modes at the outer scale of the turbulence, $\gamma_d = 4.9\times 10^{13} \, {\rm cm}^3 {\rm g}^{-1}$ is the ion-neutral drag coefficient, $C$ is the cosmic ray proton number density in the midplane, $\mu_{\rm i}$ is the atomic mass of the dominant ion species (here ${\rm C}^{+}$, so $\mu_{\rm i} = 12$), $e$ is the elementary charge and $\gamma= E_{\rm p}/m_{\rm p} c^2$ is the proton Lorentz factor. While this choice is our default expression for the diffusion coefficient, as noted above we allow users to make different choices. In practice, we accomplish this by adopting a modified expression for the streaming speed
\begin{equation}
   V_{\rm st} \approx \min\left[c,
f_\mathrm{st} V_{\rm Ai} \left(
1 + \frac{\gamma_d \; \chi \; \mathcal{M}_{\rm A} \; c \; \mu_{\mathrm{H}}^{3/2} m_\mathrm{H}^{3/2} n_\mathrm{H}^{3/2}}{4 \; \pi^{1/2} \; C \; e \; u_{\rm LA} \; \mu_{\rm i} \; \gamma^{-q+1}}
\right)
\right],
\end{equation}
where $f_\mathrm{st}$ is an arbitrary user-specified factor. We adopt $f_\mathrm{st} = 1$, corresponding to the original \citet{2020MNRAS.493.2817K} model, for all results shown in the main text, but we show results for alternative values of $f_\mathrm{st}$ in \aref{app:vai}.
The energy dependent calorimetry fraction $f_{\rm cal}\left( E_{\rm p}\right)$, which is the fraction of cosmic rays that lose their energy in hadronic collisions, in a disc geometry can be written as \citep{2020MNRAS.493.2817K}
\begin{equation}
    f_{\rm cal}\left( E_{\rm p}\right) = 1 - \left[ {}_0F_1 \left( \frac{1}{5}, \frac{16}{25}\tau_{\rm eff} \right) + \frac{3 \, \tau_{\rm eff}}{4 \, \mathcal{M}_{\rm A}^3} \; {}_0F_1 \left( \frac{9}{5}, \frac{16}{25}\tau_{\rm eff} \right) \right]^{-1},
\end{equation}
where $_{0}F_1$ is the generalised hypergeometric function and $\tau_{\rm eff}$ the dimensionless effective optical depth of the ISM, in turn, is given by
\begin{equation}
    \tau_{\rm eff} = \frac{\sigma_{\rm pp} \; \eta_{\rm pp} \; \Sigma_{\rm g} \; h_{\rm g} \; c}{2 \; D \; \mu_{\rm p} \; m_{\rm H}},
\end{equation}
where $\sigma_{\rm pp} = 40 \, {\rm mbarn}$ is the mean inelastic cross-section for proton-proton collisions, $\eta_{\rm pp} = 0.5$ is the elasticity of proton-proton collisions and $\mu_{\rm p} = 1.17$ is the number density of nucleons per proton. For a more detailed description we direct the reader to \citet{2020MNRAS.493.2817K} and to \citet{2021Natur.597..341R} for the implementation used to compute the results presented here.
The steady state spectrum for protons within the disc can be  obtained approximately by taking
\begin{equation}
    q_{\rm p} = \tau_{\rm loss} \frac{d\dot{N}_{\rm p}}{dE_{\rm p}},
\end{equation}
where $1/\tau_{\rm loss} = 1/\tau_{\rm col} + 1/\tau_{\rm diff}$ with $\tau_{\rm col} = n_{\rm H} \sigma_{\rm pp} \eta_{\rm pp} c$ and $\tau_{\rm diff} = h_{\rm g}^2/D$. We use this to to obtain $C$ by integrating $q_{\rm p}$ over all energies, using the midplane diffusion coefficient $D_{0} = V_{\rm Ai} h_{\rm g}/\mathcal{M}_{\rm A}^3$\footnote{Note that it is a reasonable approximation to adopt $V_{\rm st} = V_{\rm Ai}$ across all energies in calculating $C$ because, for physically-plausible CR spectra, the number density of CRs in the midplane is vastly dominated by CRs at lower energies where this assumption is true. \textsc{CONGRuENTS} reports the approximate steady spectrum as part of its output. However, we compute the $\gamma$-ray spectrum directly from the injection spectrum in combination with the calorimetry fraction, as described in \autoref{sec:non-thermal}. The steady state proton spectrum is not used in any further calculations. Nor do we require a solution for the protons in the halo, since we assume that CR protons in the halo do not produce any significant number of $\gamma$-rays due to the very low matter density.}.

\begin{figure}
	\includegraphics[width=\columnwidth]{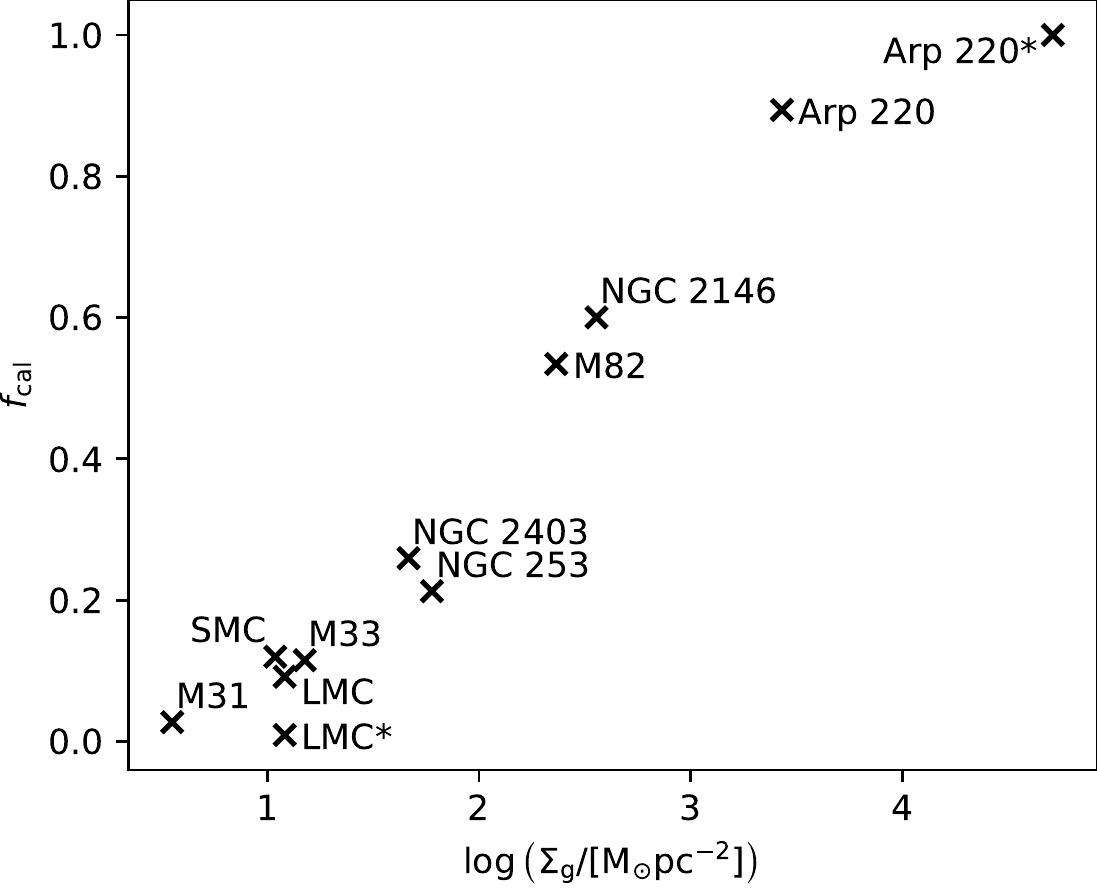}
    \caption{Sample galaxies in the $\Sigma_{\rm g}$ - $f_{\rm cal}$ plane, where $f_{\rm cal}$ is taken at the low energy limit. The starred galaxies are the `corrected' versions as discussed in \autoref{ssec:problem_galaxies}}
    \label{fig:Sigmagasfcal}
\end{figure}

The relationship between calorimetry fraction and surface gas density has been studied previously by a number of authors (e.g., \citealt{2007ApJ...654..219T}), and it is therefore interesting to see how our prescription compares to theirs. This is not entirely trivial since in our model $f_\mathrm{cal}$ depends indirectly on galaxy properties such as $h_\mathrm{g}$ and $\sigma_\mathrm{g}$ (through the diffusion coefficient $D$) as well as $\Sigma_\mathrm{g}$, so the $f_\mathrm{cal}-\Sigma_\mathrm{g}$ relationship is not one-to-one as it was in some earlier models. We can nonetheless get a sense of the relationship by evaluating both $\Sigma_{\rm g}$ and $f_{\rm cal}$ for the sample of galaxies we analyse below (see \autoref{sec:comparison}). We show this relationship for low $E_{\rm p}$ (i.e., $E_\mathrm{p}$ small enough that $V_\mathrm{st}\approx V_\mathrm{Ai}$ in \autoref{fig:Sigmagasfcal}. We see that, while the relationship is not one-to-one, there is nevertheless a very strong correlation between $f_\mathrm{cal}$ and $\Sigma_\mathrm{g}$, which is similar to that obtained by earlier authors, i.e., normal star-forming galaxies like the M31, M33, and the Magellanic Clouds are typically $\approx 10\%$ calorimetric, while the most extreme starbursts such as Arp 220 approach full calorimetry. A detailed discussion of proton calorimetry and results for some of the systems we investigate here can be found in \citet{Lacki11a}, summarised in Figure 2 of that paper.

\subsection{Cosmic ray electrons}
\label{ssec:electron_spectra}

We next describe our method for computing steady-state CR electron spectra, beginning with the overall framework (\autoref{sssec:electron_method}) and then describing our implementation of the various source and loss processes to which electrons are subject (\autoref{sssec:electron_terms}).

\subsubsection{Solution method}
\label{sssec:electron_method}

To obtain the steady-state spectra for the cosmic ray electron components (primary and secondary) in the disc and the halo, $q_{\rm e} = dN_{\rm e}/dE_{\rm e}$, we solve the kinetic equation
\begin{eqnarray}
    \lefteqn{\frac{\partial q_{\rm e}\left( E, t \right)}{\partial t} = D \nabla^2 q_{\rm e}\left( E, t \right) + Q(E,t) -
    \frac{\partial}{\partial E} \left( \dot{E} q_{\rm e}\left( E, t \right) \right) 
    }
     \\
    & & {} - q_{\rm e}\left( E, t \right) \int_{m_{\rm e} c^2}^{E} \frac{d\Gamma}{dk}\!(E) \,dk
    + \int_{E}^{\infty} q_{\rm e}\left( k, t \right) \frac{d\Gamma}{dE}\!(k)\, dk
\nonumber
\end{eqnarray}
The terms appearing on the right hand side here are, from first to last, the rate of electron diffusion in space (characterised by a spatially-independent diffusion coefficient $D$), the rate per unit time per unit energy at which sources inject CR electrons $Q$, the rate of electron transport in energy by continuous processes (where $\dot{E}$ is the rate of continuous energy loss), the rate per unit energy at which catastrophic processes cause electrons to jump from energy $E$ to a lower energy (where $d\Gamma/dk(E)$ is the differential transition rate from initial energy $E$ to final energies in the range $k$ to $k+dk$), and the rate at which catastrophic processes cause electrons to jump from some higher energy down to energy $E$. We discuss the terms that contribute to $Q$, $\dot{E}$, and $d\Gamma/dk$ below.

Our goal is to find a steady-state solution, one for which the left-hand side is zero. As in the rest of our model, we divide the problem into a disc region and a halo region, and we solve the equation separately in each. We assume that the diffusion coefficient $D$ is the same for electrons as for protons of equal rigidity, and we approximate the diffusion term $D \nabla^2 q_{\rm e}\left( E, t \right)$ as constant across the disc so that it can be written as $D \nabla^2 q_{\rm e} = q_{\rm e} D/h_{\rm g}^2$; while this is obviously a significant simplification, a more accurate treatment would require more detailed knowledge of the structure of the ISM, something that is not consistent with our goal of making a model that can be applied to large samples for which no direct measurements of gas properties are available. With this approximation, to obtain the final equation we must solve,
\begin{eqnarray}
    0 & = & \frac{\partial}{\partial E}\left(\dot{E} q_{\rm e}\right) + q_{\rm e}\frac{D}{h_{\rm g}^2} + Q
    \label{eq:electron_spectrum}
    \\
    & & {}
    + q_{\rm e} \int_{m_{\rm e} c^2}^{E} \frac{d\Gamma}{dk}\!(E)\, dk
    - \int_{E}^\infty q_{\rm e}(k) \frac{d\Gamma}{dE}\!(k)\, dk,
    \nonumber
\end{eqnarray}
where for brevity we have omitted the explicit dependence of $q_{\rm e}$, $\dot{E}$, $D$ and $Q$ on energy $E$. Solving \autoref{eq:electron_spectrum} requires some care, because it is an integro-differential equation: the final term involves an integral of $q_{\rm e}$ over all energies from $E$ to infinity. We describe our solution algorithm in \aref{app:algorithm}.

\subsubsection{Source and loss processes for cosmic ray electrons}
\label{sssec:electron_terms}

We next describe each of the processes and terms that contribute to $Q$, $\dot{E}$, and $d\Gamma/dE_{\rm f}$ in \autoref{eq:electron_spectrum}.

\paragraph{Electron sources}

In the disc, the source term $Q$ contains two contributions: $Q_1$, representing primary electrons directly accelerated by SNe, and $Q_2$, representing secondary electrons produced when primary protons produce charged pions, which subsequently decay. We neglect the contribution from tertiary electrons produced in $\gamma \gamma$ pair-production within the galaxy, since these are significantly subdominant compared to primaries and secondaries \citep{2019MNRAS.487..168P}. For primary electrons, we have $Q_1 = d\dot{N}_{\rm CR}/dE_{\rm CR}$ from \autoref{eq:cr_inject} evaluated using the values of $\Phi$ and $E_{\rm cut}$ appropriate for electrons. For secondary electrons, we adopt the spectrum of pions from \citet{2006PhRvD..74c4018K, 2009PhRvD..79c9901K}, giving
\begin{equation}
    \frac{d\dot{N}_{\pi}}{dE_\pi} = \frac{n_{\rm H} \, c}{K_{\rm \pi}} \; \beta \; \sigma_{\rm pp}(E_{\rm p}) \; \frac{d\dot{N}_{\rm p}}{dE_{\rm p}} f_{\rm cal}\left(E_{\rm p}\right),
    \label{eq:pion_energies}
\end{equation}
and then compute the resulting electron injection spectrum as
\begin{equation}
    Q_2 = \frac{d\dot{N}_{\rm e}}{dE_{\rm e}} = \int_{E_{\rm e}/E_{\pi}}^{1} f\left( x \right) \frac{d\dot{N}_{\pi}}{dE_\pi}\left( \frac{E_{\rm e}}{x} \right) \frac{dx}{x}.
    \label{eq:secondary_electrons}
\end{equation}
Here $d\dot{N}_{\rm p}/dE_p$ and $f_{\rm cal}(E_{\rm p})$ are the CR proton injection rate and calorimetry fraction evaluated as described in \autoref{ssec:proton_spectra}, $x = E_e/E_\pi$ (so $d\dot{N}_\pi/dE_\pi$ is evaluated at pion energy $E_\pi = E_e/x$), and $f(x) = 2 f_{\nu_\mu^{(2)}}$ is the fitting function for secondary electron injection taken from \citet{2006PhRvD..74c4018K, 2009PhRvD..79c9901K}. Thus our total injection rate in the disc is $Q_{\rm disc} = Q_1 + Q_2$. By contrast, in the halo we have no local sources of CR electrons; instead, the only source is electrons diffusing out of the disc. The rate at which this process provides electrons simply follows from our adopted form of the diffusion term,
\begin{equation}
    Q_{\rm halo} = \frac{D}{h_{\rm g}^2} q_{\rm e, disc},
\end{equation}
where $q_{\rm e, disc}$ is the solution we obtain for \autoref{eq:electron_spectrum} in the disc.

\paragraph{Continuous losses: synchrotron emission and ionisation.}

The term $\dot{E}$ in \autoref{eq:electron_spectrum} represents continuous losses -- those where the change in electron energy per interaction is small enough that it can be approximated as infinitesimal. The two loss mechanisms to which this approximation applies are synchrotron emission and ionisation / Coulomb losses. For the former, we adopt the standard expression from \citet{1970RvMP...42..237B},
\begin{equation}
    \dot{E}_{\rm sync} = - \frac{4}{3} \sigma_{\rm T} c \left(\frac{E_{\rm e} \beta}{m_{\rm e} c^2}\right)^2 \frac{B^2}{8 \pi}
\end{equation}
where $\sigma_{\rm T}$ is the Thomson cross section and $\beta$ the reduced velocity of the electron. For the latter, we adopt the model by \citet{2002cra..book.....S} for an ISM composed of 91\% H and 9\% He by number with average electronic excitation energies $\Delta E_{\rm H} = 15$ eV and $\Delta E_{\rm He} = 41.5$ eV. For the disc, which is predominantly neutral, we have
\begin{equation}
\begin{split}
    \dot{E}_{\rm ion} = &-\frac{9}{4} c \sigma_{\rm T} m_{\rm e} c^2 n_{\rm H} \mu_{\rm ISM} \left[ \ln\left( \frac{E_{\rm e}}{m_{\rm e} c^2} \right) + 0.91 \frac{2}{3} \ln\left( \frac{m_{\rm e} c^2}{\Delta E_{\rm H}} \right) \right.\\ &+ \left. 0.09 \frac{4}{3} \ln\left( \frac{m_{\rm e} c^2}{\Delta E_{\rm He}} \right) \right],
    \label{eq:edot_ion}
\end{split}
\end{equation}
where $\mu_{\rm ISM}$ is the ratio of the total to the hydrogen number density of ISM constituents and $n_{\rm H}$ is the midplane density of H nuclei. For the halo, where the gas is ionised and Coulomb losses dominate, we have instead
\begin{equation}
    \dot{E}_{\rm Coul} = - \frac{3}{4} c \sigma_{\rm T} m_{\rm e} c^2 n_{\rm H} \mu_{\rm ISM} \left[ \ln\left( \frac{E_{\rm e}}{m_{\rm e} c^2} \right)  + 2 \ln\left( \frac{m_{\rm e} c^2}{h \nu_{\rm p}} \right) \right],
\end{equation}
where $\nu_{\rm p} = e \sqrt{ n_{\rm H} \mu_{\rm ISM} / \pi m_{\rm e} }$ is the plasma frequency. Thus our final continuous loss rates are $\dot{E} = \dot{E}_{\rm sync} + \dot{E}_{\rm ion}$ in the disc, and $\dot{E} = \dot{E}_{\rm sync} + \dot{E}_{\rm Coul}$ in the halo.

\paragraph{Catastrophic losses: inverse Compton and bremsstrahlung.}

Our terms $d\Gamma/dE_{\rm f}$ in \autoref{eq:electron_spectrum} represent processes that can cause large jumps in the electron energy. The two processes of this type that we include are inverse Compton scattering\footnote{In fact inverse Compton scattering is catastrophic only in the Klein-Nishina regime where the initial electron and photon energies $E_\mathrm{i}$ and $E_\mathrm{ph}$ obey $\sqrt{E_\mathrm{i} E_\mathrm{ph}} \gg m_e c^2$. In the opposite limit, the Thomson regime, it is well-approximated as continuous. We choose to treat inverse Compton scattering as catastrophic in all cases because doing so avoids the need to impose an arbitrary switch between them. The treatment we adopt below correctly recovers the Thomson regime limit where it should.} and bremsstrahlung. For the latter, if we ignore recoil of the nucleus, the final electron energy $k$ and initial electron energy $E$ are related by $k = E - E_\gamma$, where $E_\gamma$ is the energy of the emitted photon. We therefore have
\begin{equation}
\frac{d\Gamma}{dk}\!(E) = \left.\frac{d\dot{N}_{\gamma}}{dE_{\gamma}}\right|_{E_\gamma = E-k},
\end{equation}
where $d\dot{N}_\gamma/dE_\gamma$ is the rate per unit time per unit photon energy at which photons are emitted, evaluated at photon energy $E_{\gamma} = E-k$. We compute the distribution of photon energies from the differential cross section given by \citet{1970RvMP...42..237B}
\begin{equation}
\frac{d\sigma_{\rm bs}}{dE_{\gamma}} = \frac{3}{8 \pi} \sigma_{\rm T} \alpha E_{\gamma}^{-1} \left\{ \left[ 1 + \left( 1 - \frac{E_{\gamma}}{E_{\rm i}} \right)^{2} \right] \Phi_{1} - \frac{2}{3} \left( 1 - \frac{E_{\gamma}}{E_{\rm i}} \right) \Phi_{2} \right\},
\label{eq:sigma_bs}
\end{equation}
where $\alpha$ is the fine structure constant and the functions $\Phi_{1}$ and $\Phi_{2}$ for hydrogen in the shielded regime are given in Table II in \citet{1970RvMP...42..237B}. For an ISM that is mostly in the form of atomic hydrogen the shielded regime is the appropriate choice. Thus our final expression for the photon production rate (and thus the catastrophic loss rate) due to bremsstrahlung is
\begin{equation}
\frac{d\dot{N}_{\gamma}}{dE_{\gamma}} = c n_{\rm H} \frac{d\sigma_{\rm bs}}{dE_{\gamma}}
\label{eq:bremsstrahlung}
\end{equation}
where $n_{\rm H}$ is the hydrogen number density of the medium through which the CRs propagate. In principle we should use the unshielded bremsstrahlung cross section in place of \autoref{eq:sigma_bs} in the halo, since the gas is fully ionised, but in practice the halo density is so low that bremsstrahlung is unimportant compared to synchrotron and inverse Compton losses, and thus we do not bother to make this correction.

For inverse Compton scattering the relationship between $E_{\rm i}$, $E_{\rm f}$, and $E_\gamma$ is somewhat more complex, since the initial photon that is upscattered by the electron also carries a non-zero energy $E_{\rm ph}$ that cannot be neglected because, in the Thomson limit, weak scattering events for which $E_\gamma/E_{\rm ph}-1 \ll 1$ contribute significantly to the total energy loss rate. By conservation of energy we have $E_{\rm i} + E_{\rm ph} = E_{\rm f} + E_{\gamma}$. Consider CR electrons of energy $E_{\rm i}$ scattering photons with energies in the range $E_{\rm ph}$ to $E_{\rm ph} + dE_{\rm ph}$, and let $dn_{\rm ph}$ be the number density of such photons. In this case we can write the rate per unit time per unit energy at which electrons are scattered to energy $E_{\rm f}$ as
\begin{equation}
    \frac{d\Gamma}{dk}\!(E) = dn_{\rm ph} \left.\frac{d\dot{N}_\gamma}{dE_\gamma}\right|_{E, E_{\rm ph},  E_\gamma = E + E_{\rm ph} - k},
    \label{eq:IC_scatter_differential}
\end{equation}
where the term on the right hand side is the rate per unit time per unit scattered photon energy that an electron of energy $E$ scatters photons of initial energy $E_{\rm ph}$ to final photon energy $E_\gamma = E + E_{\rm ph} - k$. For an isotropic field of photons to be scattered, we can write this scattering rate as \citep{1968PhRv..167.1159J,1970RvMP...42..237B}:
\begin{equation}
\frac{d\dot{N}_{\gamma}}{dE_{\gamma}} = \frac{3}{4} \sigma_{\rm T}c \frac{m_{\rm e}^2 c^4}{E_{\rm i}^2 E_{\rm ph}} G\left( q, \Gamma_{\rm e} \right),
\end{equation}
where $q = E_{\gamma} / \Gamma_{\rm e} \left( E_{\rm i} - E_{\gamma} \right)$ and $\Gamma_{\rm e} = 4 E_{\rm ph} E_{\rm i} / m_{\rm e}^2 c^4 $. The function $G\left( q, \Gamma_{\rm e} \right)$ encodes the Klein-Nishina cross-section and is given as
\begin{equation}
G\left( q, \Gamma_{\rm e} \right) = 2 q \ln{q} \left( 1 + 2 q \right) \left( 1 - q \right) + \frac{1}{2} \frac{\left( \Gamma_{\rm e} q \right)^2}{1+\Gamma_{\rm e} q} \left( 1 - q \right).
\end{equation}
To evaluate the total catastrophic transition rate that appears on the left hand side of \autoref{eq:IC_scatter_differential}, we must integrate over the distribution of photon energy $dn_{\rm ph}/dE_{\rm ph}$ present in the ISRF being scattered, evaluating the cross section at photon energy $E_\gamma$ such that the final electron energy $k$ is held constant. Doing so gives
\begin{eqnarray}
    \lefteqn{\frac{d\Gamma}{dk}\!(E) = \frac{3}{4}\sigma_{\rm T} c \frac{m_e^2 c^4}{E^2} \times {}
    }
    \label{eq:IC_scatter_total}
    \\
    & & \quad \int_{E_{\rm ph,min}}^{E_{\rm ph,max}} \frac{dn_{\rm ph}}{dE_{\rm ph}} \frac{\left.G(q,\Gamma_e)\right|_{E_\gamma = E+E_{\rm ph}-k}}{E_{\rm ph}} dE_{\rm ph}.
    \nonumber
\end{eqnarray}
Here the function $G(q,\Gamma_e)$ is evaluated at the initial photon energy $E_{\rm ph}$ and final photon energy $E_\gamma$ as indicated, and the limits of integration $E_{\rm ph,min}$ and $E_{\rm ph,max}$ correspond to the kinematic limits on the minimum and maximum photon energies that can give rise to a scattering in which the electron energy changes from $E$ to $k$. These limits correspond to $q$ being in the range $m_{\rm e}^2 c^4/\left( 4 E \right) < q \leq 1$, which from the definition of $q$ implies
\begin{equation}
    \begin{split}
    &E_{\rm ph, min} = \frac{\left( E - k\right) m_{\rm e}^2 c^4}{4 E k - m_{\rm e}^2 c^4}\\
    &E_{\rm ph, max} = \frac{1}{2} \left[ \left( k - \frac{m_{\rm e}^2 c^4}{4 E} \right) + \sqrt{\left( k + \frac{m_{\rm e}^2 c^4}{4 E} \right)^2 - m_{\rm e}^2 c^4}\right].
    \end{split}
\end{equation}
Below we shall also require the spectrum of scattered photons, which we can again obtain by integrating \autoref{eq:IC_scatter_differential} over $E_{\rm ph}$, but this time at constant $E_{\gamma}$ rather than constant $k$ as in \autoref{eq:IC_scatter_total}; the resulting expression is
\begin{equation}
\label{eq:inverseCompton}
\left.\frac{d\dot{N}_{\gamma}}{dE_{\gamma}}\right|_{E} = \frac{3}{4} \sigma_{\rm T}c \frac{m_{\rm e}^2 c^4}{E^2} \int_{E_{\rm ph,min}}^{E_{\rm ph,max}} \frac{dn_{\rm ph}}{dE_{\rm ph}} \frac{G\left( q, \Gamma_{\rm e} \right)}{E_{\rm ph}} dE_{\rm ph},
\end{equation}
with limits
\begin{equation}
    \begin{split}
    &E_{\rm ph, min} = \frac{E_{\gamma} m_{\rm e}^2 c^4}{4 E \left(E - E_{\gamma}\right)},\\
    &E_{\rm ph, max} = \frac{E_{\gamma} E}{E - E_{\gamma}}.
    \end{split}
\end{equation}

\section{Non-thermal emission}
\label{sec:non-thermal}

Once we have determined the steady-state CR spectra, the next step in our modeling procedure is to calculate the non-thermal emission they produce. Our model includes both hadronic $\gamma$-rays and neutrinos (\autoref{ssec:hadronic}) and leptonic $\gamma$-rays and radio emission (\autoref{ssec:leptonic}), and also includes the effects of free-free emission and absorption (\autoref{ssec:free-free}), which are important for the radio spectra in some systems.

\subsection{Hadronic emission}
\label{ssec:hadronic}

We calculate the hadronic $\gamma$-ray spectrum using the same simple model we use for the steady state proton spectrum, whereby we assume that the only two loss mechanisms of relevance are diffusive escape and collisional pion production. Making the same assumptions as in \autoref{ssec:proton_spectra}, the $\gamma$-ray emission rate per unit time per unit energy is
\begin{equation}
    \phi_\gamma \equiv \frac{d\dot{N}_{\gamma}}{dE_{\gamma}} = \int_{m_{\rm p} c^2}^{\infty} \left[ \frac{1}{\sigma_{\rm pp}} \frac{d\sigma_{\gamma}}{dE_{\gamma}} \right] f_{\rm cal} \frac{d\dot{N}_{\rm p}}{dE_{\rm p}}dE_{\rm p}.
\end{equation}
Here $f_{\rm cal}$ is the calorimetry fraction as a function of proton energy $E_{\rm p}$ (c.f.~\autoref{ssec:proton_spectra}), $d\dot{N}_{\rm p}/dE_{\rm p}$ is the energy-dependent proton injection rate by SNe, $d\sigma_{\gamma}/dE_{\gamma}$ is the differential $\gamma$-ray production cross-section as a function of proton energy taken from \citet{2014PhRvD..90l3014K}, and $\sigma_{\rm pp} \approx 40 \ {\rm mbarn}$ is the mean inelastic cross-section for proton-proton collisions.

We calculate hadronic neutrino emission following the approach of \citet{2006PhRvD..74c4018K, 2009PhRvD..79c9901K}. Once initial CR proton collisions produce charged pions (at a rate per unit energy $d\dot{N}_\pi/dE_\pi$ given by \autoref{eq:pion_energies}), neutrinos are emitted in a two stage process. In the first stage, the pions decay to muons and muon neutrinos, and in the second the muon decays to an electron, a muon neutrino, and an electron neutrino (where we do not distinguish neutrino from anti-neutrino). The neutrino production rates as a function of energy for both stages, and for both neutrinos during the second stage, can be expressed approximately using the same functional form we use for the distribution of secondary electron energies (\autoref{eq:secondary_electrons}), i.e.,
\begin{equation}
    \frac{d\dot{N}_{\nu}}{dE_{\nu}}= \int_{E_{\rm \nu}/E_{\pi}}^{1} f(x) \frac{d\dot{N}_{\pi}}{dE_\pi}\left( \frac{E_{\nu}}{x} \right) \frac{dx}{x},
    \label{eq:neutrino_spectrum}
\end{equation}
where $x = E_\nu/E_\pi$ (and thus $d\dot{N}_\pi/dE_\pi$ is evaluated at $E_\pi = E_\nu/x$). Only the fitting functions $f(x)$ differ between the channels. For the muon emission during the first stage we have $f(x) = f_{\nu_\mu^{(1)}}(x) = 2/\lambda$ for $x < \lambda$ and $f(x) = 0$ for $x > \lambda$, where $\lambda = 1-(m_{\mu}/m_{\pi})^2$, and $m_\mu$ and $m_\pi$ are the muon and (charged) pion rest mass, respectively. For the second stage, the fitting functions for the electron and muon neutrinos are given by \citeauthor{2006PhRvD..74c4018K}'s functions $2f_{\nu_e}(x)$ and $2f_{\nu_\mu^{(2)}}$, respectively. The total emitted neutrino spectrum over all neutrino flavours is the sum of these contributions, i.e., \autoref{eq:neutrino_spectrum} evaluated using $f(x) = 2[f_{\nu_\mu^{(1)}}(x) + f_{\nu_\mu^{(2)}} + f_{\nu_e}(x)]$.

\subsection{Leptonic emission}
\label{ssec:leptonic}

To obtain the total emitted spectra for CR leptons $d\dot{N}_\gamma/dE_{\gamma}$ we integrate the photon production rate per electron $(d\dot{N}_\gamma/dE_{\gamma})_i$ from all emission processes $i$ -- bremsstrahlung, inverse Compton, and synchrotron emission -- over the CR electron energy distribution $q_{\rm e}$:
\begin{equation}
\phi_\gamma = \sum_{i} \int_{m_{\rm e} c^2}^{\infty} \left(\frac{d\dot{N}_{\gamma}}{dE_\gamma}\right)_i q_{\rm e}\, dE_{\rm e}.
\label{eq:emit_lepton}
\end{equation}
For bremsstrahlung and inverse Compton emission, we have already provided expressions for the relevant photon production rates -- these are given by \autoref{eq:bremsstrahlung} and \autoref{eq:IC_scatter_differential}, respectively. For synchrotron emission we have
\begin{equation}
\label{eq:specsync}
\left(\frac{d\dot{N}_{\gamma}}{dE_\gamma}\right)_{\rm sync} = \frac{\pi \sqrt{3} e^3 B}{2 h m_e c^2} \frac{ \mathcal{F}\left( E_{\gamma}/E_{\rm \gamma, c}\right)}{E_{\rm \gamma, c}},
\end{equation}
where the function $\mathcal{F}(E_\gamma/E_{\rm \gamma, c})$ is the integral over the modified Bessel function of order 5/3, and $E_{\gamma,\rm c} = 3 e B E_{\rm e}^2 h/\left( 2 \pi^2 m_{\rm e}^3 c^5\right)$ is the energy corresponding to the synchrotron critical frequency.

\subsection{Free-free absorption and emission}
\label{ssec:free-free}

Free-free absorption by ionised gas can be a significant source of opacity at higher radio wavelengths and is therefore important to consider when predicting galaxies' non-thermal radio continuum emission. One common approach to estimating this effect is to treat this absorption as a uniform screen with an optical depth determined by an average ionisation fraction for the ISM \citep[e.g.,][]{2022A&A...657A..49K}. However, the geometry implicitly assumed in this approach is unrealistic: in reality, the ISM is segregated into discrete phases, with the ionised phase existing mainly in regions of near complete ionisation around ionising sources and the intervening space filled mostly by material with very low ionisation fraction. Compared to a uniform screen this geometry leads to a lower covering fraction of absorption, but also vastly higher opacity within each ionised region, since the free-free opacity scales as the square of density. Due to this effect, we find that the uniform screen approximation underestimates the free-free optical depth somewhat. For this reason, we employ a more detailed procedure, explained below, that assumes that fully ionised H~\textsc{ii} regions are the main source of opacity.

On average star formation produces an ionising photon budget per unit mass of stars formed of approximately $\Psi = 4.2 \times 10^{60} M_{\odot}^{-1}$ \citep{2017stfo.book.....K}. We assume that these photons ionise Str\"{o}mgren spheres of characteristic radius $r_{\rm s}$. Within such a sphere the total recombination rate is then $\mathcal{R} = \left(4/3\right) \pi r_{\rm s}^3 \alpha_{\rm B} f_{\rm e} n_{\rm i}^2$, where $n_{\rm i}$ is the number density of H nuclei in the ionised region, and $f_{\rm e}$ is the number of free electrons per hydrogen nucleon; assuming He is singly ionised, as is the case for over the bulk of the volume of H~\textsc{ii} regions, $f_{\rm e} = 1.1$. The hydrogen recombination rate for case B $\alpha_{\rm B}$ is given by
(this and following expressions from \citealt{2011piim.book.....D}):
\begin{equation}
\alpha_{\rm B} = 2.54 \times 10^{-13} T_4^{-0.8163-0.0208 \ln T_4} \mbox{ cm}^3\mbox{ s}^{-1}
\end{equation}
where $T_4$ is the temperature $T$ of the ionised gas divided by $10^4$ K; we adopt $T_4 = 1$. The coefficient for free-free absorption by the ionised gas, again assuming that He is singly ionised and thus that the mean ion charge $Z_{\rm i} \approx 1$, is
\begin{equation}
\kappa_{\nu, \rm ff} = \frac{4 e^6}{3 c h} \left( \frac{2 \pi}{3 m_{\rm e}^3 k_{\rm B} T} \right)^{1/2} \frac{\left( 1 - {\rm e}^{-h \nu / k_{\rm B} T} \right)}{\nu^3} f_{\rm e} n_{\rm i}^2 g_{\rm ff, i}
\end{equation}
where $g_{\nu, \rm ff}$ is the Gaunt factor. We compute this using a piecewise approximation taken from \citet{2011piim.book.....D}
\begin{equation}
g_{\nu, \rm ff} = \begin{cases} 
4.691 \left[1 - 0.118 \ln \left(\frac{\nu_9}{10 T_4^{3/2}}\right)\right],
& \nu_9 < 1
\\
\ln\left\{
\exp\left[
5.960 - \frac{\sqrt{3}}{\pi} \ln \left(\nu_9 T_4^{-3/2}\right)
\right] + {\rm e}
\right\},
& \nu_9 > 1,
\end{cases}
\end{equation}
where $\nu_9 = \nu/10^9$ Hz. The optical depth across a single Str\"omgren sphere is
\begin{equation}
    \tau_{\nu,\rm ff} \approx \kappa_{\nu, \rm ff} r_{\rm s},
\end{equation}
where we have omitted geometric factors of order unity that depend on the exact path length through the H~\textsc{ii} region.

Now consider a segment of a galactic disc with a star formation rate per unit area $\dot{\Sigma}_*$, and thus an ionising photon production rate per unit area $\dot{\Sigma}_* \Psi$. From ionisation balance, the number of Str\"omgren spheres per unit area must therefore be $\dot{\Sigma}_* \Psi/\mathcal{R}$, and the fraction of the galactic disc covered by Str\"omgren spheres is thus $\pi r_{\rm s}^2 \dot{\Sigma}_* \Psi/\mathcal{R}$. Now consider a scenario where a photon is emitted close to the galactic mid plane and then intersects some number of H~\textsc{ii} regions on its way out of the galaxy. Since the photon will traverse on average half the thickness of the disc, the expected number of Str\"omgren spheres through which it passes before escaping is 
\begin{equation}
\langle N \rangle = \frac{1}{2} \frac{\dot{\Sigma}_{\ast} \Psi}{\mathcal{R}} \pi r_{\rm s}^2 = \frac{3 \dot{\Sigma}_{\ast} \Psi}{8 r_{\rm s} \alpha_{\rm B} f_{\rm e} n_{\rm i}^2} = \frac{3 \dot{\Sigma}_{\ast} \Psi \kappa_{\nu}}{8 \tau_{\rm \nu, ff} \alpha_{\rm B} f_{\rm e} n_{\rm i}^2}
\end{equation}
Note that, contrary to appearances, this expression does not contain any explicit dependence on the ionised gas density $n_{\rm i}$, since $\kappa_{\nu, \rm ff} \propto n_{\rm i}^2$. All the dependence on ionised gas density is implicitly contained in the optical depth $\tau_{\nu,\rm ff}$. Assuming, for the moment, that the ionisation regions are uniformly distributed in the ISM and that non-thermal photons are launched from regions uncorrelated with these ionisation regions, the probability that any such photon trajectory intersects $N$ of the spheres, is given by a Poisson distribution:
\begin{equation}
P\left(N\right) = \frac{\langle N \rangle^N e^{-\langle N \rangle}}{N!} \, .
\end{equation}
The mean transmittance is obtained by summing over all lines of sight
\begin{equation}
\langle f_{\rm trans} \rangle = \sum_{N=0}^{\infty} P\left(N\right) e^{-N \tau_{\rm \nu, ff}}\; ,
\end{equation}
an expression that can be evaluated analytically to yield 
\begin{equation}
\langle f_{\rm trans} \rangle = \exp\left[ \langle N \rangle \left( e^{-\tau_{\rm \nu, ff}} - 1 \right) \right]
\end{equation}

We now introduce the effective mean optical depth $\langle \tau_{\rm \nu, ff} \rangle = -\ln \langle f_{\rm trans} \rangle$, which evaluates to
\begin{equation}
\langle \tau_{\rm \nu, ff} \rangle = \langle N \rangle \left( 1 - e^{-\tau_{\rm \nu, ff}} \right) =\frac{3 \dot{\Sigma}_{\ast} \Psi \kappa_{\nu}}{8 \alpha_{\rm B} f_{\rm e} n_{\rm i}^2} \left(\frac{1 - e^{-\tau_{\rm \nu, ff}}}{\tau_{\rm \nu, ff}}\right)
\label{eq:tau_ff_mean}
\end{equation}
The non-thermal radio emission is attenuated by a factor $\exp(-\langle\tau_{\nu,\rm ff}\rangle)$. For consistency we must also include the free-free emission itself in our predicted spectra, since the absorption of the non-thermal radiation will be filled in by thermal free-free emission; per Kirchoff's law, in the limit of high $\langle\tau_{\nu,\rm ff}\rangle$, the observed photon flux must approach that of a blackbody at temperature $T$. Thus our final expression for the photon spectrum escaping the galaxy is
\begin{equation}
    \phi_\gamma = e^{-\langle\tau_{\nu,\rm ff}\rangle} \phi_{\gamma,\rm emit} + 
    \left(1 - e^{-\langle\tau_{\nu,\rm ff}\rangle}\right) 4\pi R_{\rm e}^2 c \phi_{\gamma,\rm BB,10^4\,K},
\end{equation}
where $\phi_{\gamma,\rm emit}$ is the intrinsic emitted spectrum (computed from \autoref{eq:emit_lepton}) and $\phi_{\gamma,\rm BB,10^4\,K}$ is the photon energy distribution for a blackbody radiation field at a temperature of $10^4$ K.

The final step is to evaluate $\langle\tau_{\nu,\rm ff}\rangle$, which depends only on physical constants, on the star formation rate per unit area $\dot{\Sigma}_*$ (which is known for each galaxy), and on $\tau_{\nu,\rm ff}$ the optical depth of a single H~\textsc{ii} region, which is not. However, we note that the term in parentheses on the right hand side of \autoref{eq:tau_ff_mean}, which contains the dependence on $\tau_{\nu,\rm ff}$, is strictly $< 1$, and is close to unity if $\tau_{\rm\nu,ff} \lesssim 1$. We are most concerned with free-free absorption at frequencies $\nu_9 \sim 1$, and observed individual H~\textsc{ii} regions generally have optical depths of at most unity at this frequency \citep{2011piim.book.....D}; we therefore adopt the small $\tau_{\rm \nu,ff}$ limit and set the factor in parentheses to unity.

\section{Application to nearby galaxies}
\label{sec:comparison}

\begin{figure}
	\includegraphics[width=\columnwidth]{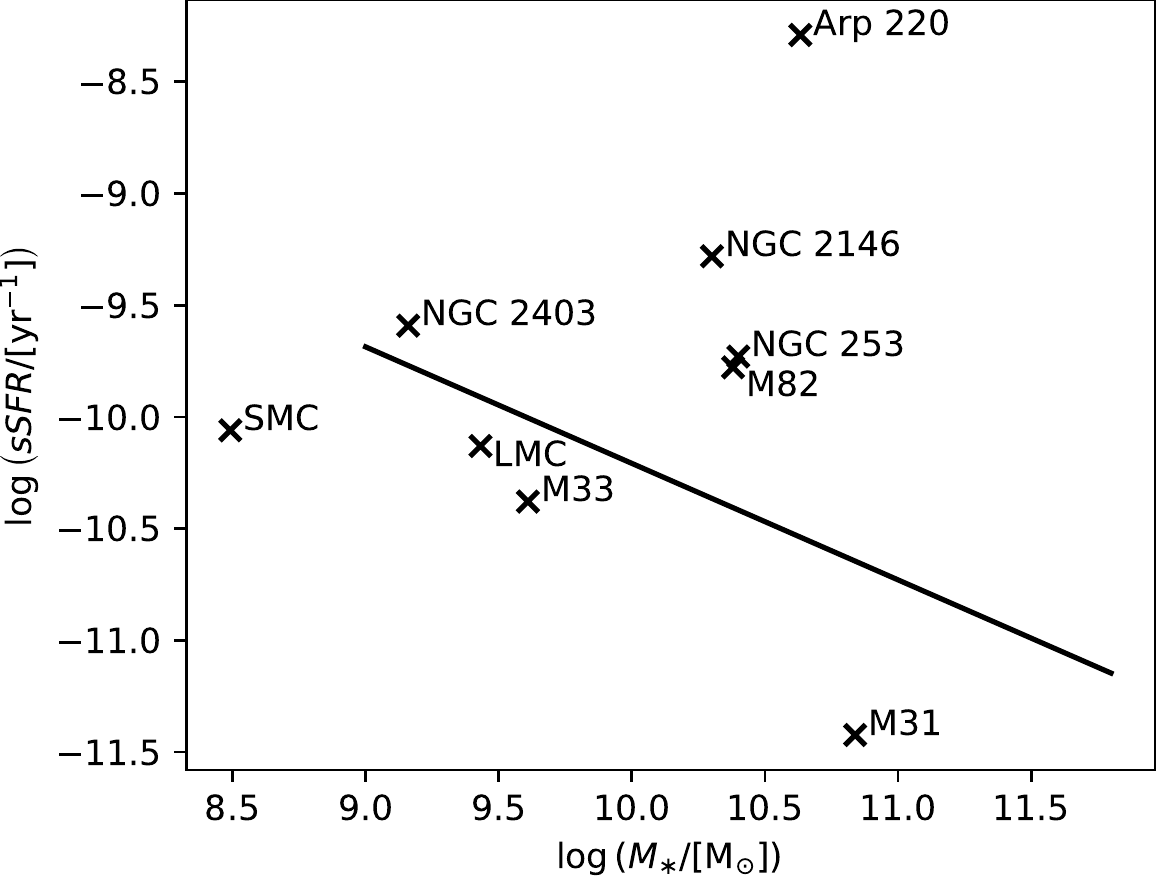}
    \caption{Sample galaxies in relation to the star-forming main sequence. The black line is the star-forming sequence obtained by \citet{2014ApJS..214...15S} and \citet{2018ApJ...853..179T} at $z = 0$.}
    \label{fig:SFMS}
\end{figure}

We have now specified the full procedure by which \textsc{CONGRuENTS} predicts the steady state CR spectra (\autoref{sec:cr_spectra}) and non-thermal emission (\autoref{sec:non-thermal}) of galaxies from the minimal set of observables to which we have access. Our next step is to apply our model to nearby galaxies for which observations of the non-thermal spectrum are available. This serves the dual purpose of validating our  methodology (\autoref{ssec:good_galaxies}) and identifying areas where (inevitably given the extraordinarily limited range of observational inputs) our method encounters problems (\autoref{ssec:problem_galaxies}). For this purpose, we choose a number of nearby radio- and $\gamma$-ray-observed star-forming galaxies, excluding any that are suspected to have a significant contribution from an active galactic nucleus. This sample spans three orders of magnitude in star formation rate, from the nearly-quenched galaxy M31 to the ultra-luminous infrared galaxy Arp 220 that is undergoing  a nuclear starburst. The measured input parameters we have used in the calculation of our results are given in \autoref{tab:gals}. We also show where our sample galaxies sit relative to the star-forming main sequence at $z=0$ \citep{2014ApJS..214...15S,2018ApJ...853..179T} in \autoref{fig:SFMS}; the figure makes it clear that our sample offers good coverage in terms of both stellar mass and specific star formation rate.
Note that we not include the Milky Way as one of our comparison galaxies due to the difficulty of obtaining a total integrated spectrum for it from our vantage point within the Galactic plane. Any such comparison would be dominated by the significant uncertainties on the observed quantities, and therefore not be useful for the purposes of constraining the model.

\begin{table*}
\centering
\begin{tabular}{l *{5}{c}}
\hline \hline
Galaxy & $D \, \left[\rm Mpc \right]$ & $M_{\ast} \, \left[\rm 10^{10} \, M_{\odot} \right]$ & $R_{\rm e} \, \left[\rm kpc \right]$ & $\dot{M}_{\ast} \, \left[\rm M_{\odot} \, yr^{-1} \right]$ & $\log\left(\mbox{sSFR}/\left[\rm yr^{-1} \right]\right)$\\
\hline
Arp 220 & $80.9^{\left(1\right)}$ & $4.3^{\left(2\right)}$ & $3.4^{\left(3\right)}$ & $220^{\left(1\right)}$ & $-8.29$\\
NGC 2146 & $17.2^{\left(1\right)}$ & $2.0^{\left(4\right)}$ & $1.8^{\left(5\right)}$ & $10.5^{\left(4\right)}$ & $-9.28$\\
NGC 2403 & $3.18^{\left(1\right)}$ & $0.14^{\left(6\right)}$ & $1.7^{\left(7\right)}$ & $0.37^{\left(1\right)}$ & $-9.58$\\
M82 & $3.53^{\left(1\right)}$ & $2.4^{\left(8\right)}$ & $1.0^{\left(7\right)}$ & $4.0^{\left(16\right)}$ & $-9.66$\\
NGC 253 & $3.56^{\left(1\right)}$ & $2.5^{\left(9\right)}$ & $4.2^{\left(7\right)}$ & $4.68^{\left(9\right)}$ & $-9.73$\\
LMC & $0.050^{\left(1\right)}$ & $0.27^{\left(14\right)}$ & $2.63^{\left(13\right)}$ & $0.2^{\left(1\right)}$ & $-10.1$\\
SMC & $0.060^{\left(1\right)}$ & $0.031^{\left(14\right)}$ & $1.15^{\left(13\right)}$ & $0.027^{\left(1\right)}$ & $-10.1$\\
M33 & $0.91^{\left(1\right)}$ & $0.41^{\left(15\right)}$ & $1.51^{\left(7\right)}$ & $0.17^{\left(12\right)}$ & $-10.4$\\
M31 & $0.77^{\left(1\right)}$ & $6.92^{\left(11\right)}$ & $2.46^{\left(11\right)}$ & $0.26^{\left(1\right)}$ & $-11.4$\\
\hline \hline
\end{tabular}
\caption{Galaxy properties used in our calculations; from left to right, these are the distance $D$, the stellar mass $M_*$, the effective radius $R_{\rm e}$, and the star formation rate $\dot{M}_*$. For convenience we also report the specific star formation rate $\mbox{sSFR} = \dot{M}_*/M_*.$ Values are taken from the following sources:$\left(1\right)$ \citet{2020A&A...641A.147K}, $\left(2\right)$ \citet{2008MNRAS.384..875R}, $\left(3\right)$ \citet{1990NASCP3098..321W}, $\left(4\right)$ \citet{2011ApJ...738...89S}, $\left(5\right)$ \citet{2018PASJ...70...49S}, $\left(6\right)$ \citet{2018MNRAS.474.4366P}, $\left(7\right)$ \citet{2003AJ....125..525J}, $\left(8\right)$ \citet{2012MNRAS.419.2095M}, $\left(9\right)$ \citet{2018ApJ...864...40P}, $\left(10\right)$ \citet{2016MNRAS.456.4128R}, $\left(11\right)$ \citet{1976PASJ...28...27I}, $\left(12\right)$ \citet{2018MNRAS.479..297W}, $\left(13\right)$ \citet{1960ApJ...131..574D}, $\left(14\right)$ \citet{2009IAUS..256...81V}, $\left(15\right)$ \citet{2019ApJS..245...25J}, $\left(16\right)$ \citet{2009ApJ...697.2030S}.}
\label{tab:gals}
\end{table*}

\subsection{Observational comparison}
\label{ssec:good_galaxies}

\begin{figure*}
	\includegraphics[width=\textwidth]{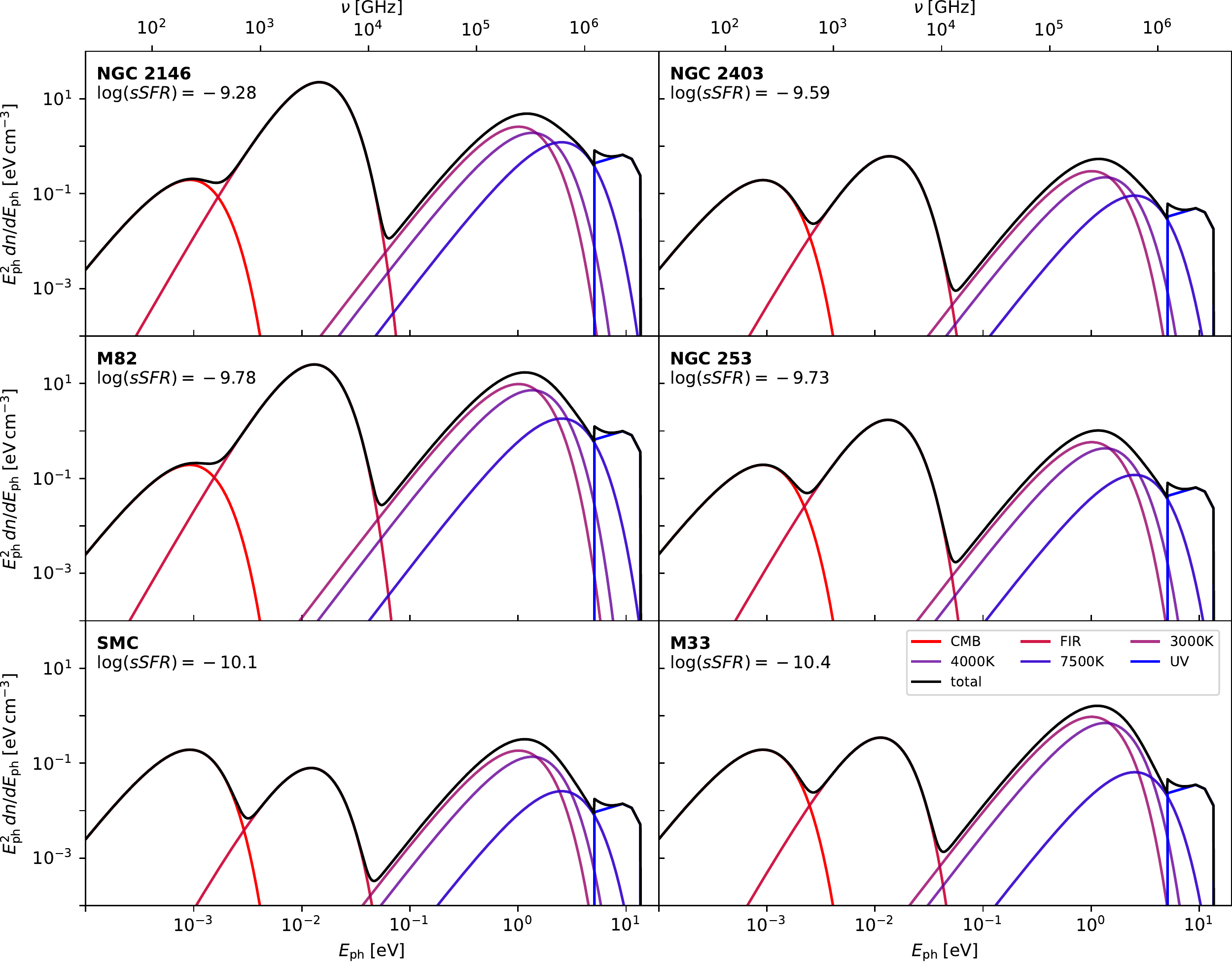}
    \caption{Model radiation fields for galaxies as determined using the method described in \autoref{ssec:isrf}. The black line shows the total ISRF, while red through blue lines show the individual components that make it up; these are, from left the right, the CMB, the dust-reprocessed radiation field, the 3000 K, 4000 K, and 7500 K blackbody starlight components, and the FUV component. For each galaxy we indicate the specific star formation rate sSFR in units of ${\rm yr^{-1}}$. The high energy cutoff of the UV component corresponds to the $13.6 {\rm eV}$ ionisation threshold of hydrogen, whereas the low energy cutoff is not physical but a result of the truncated parametrisation of the UV spectrum.}
    \label{fig:radfields}
\end{figure*}

\begin{figure*}
	\includegraphics[width=\textwidth]{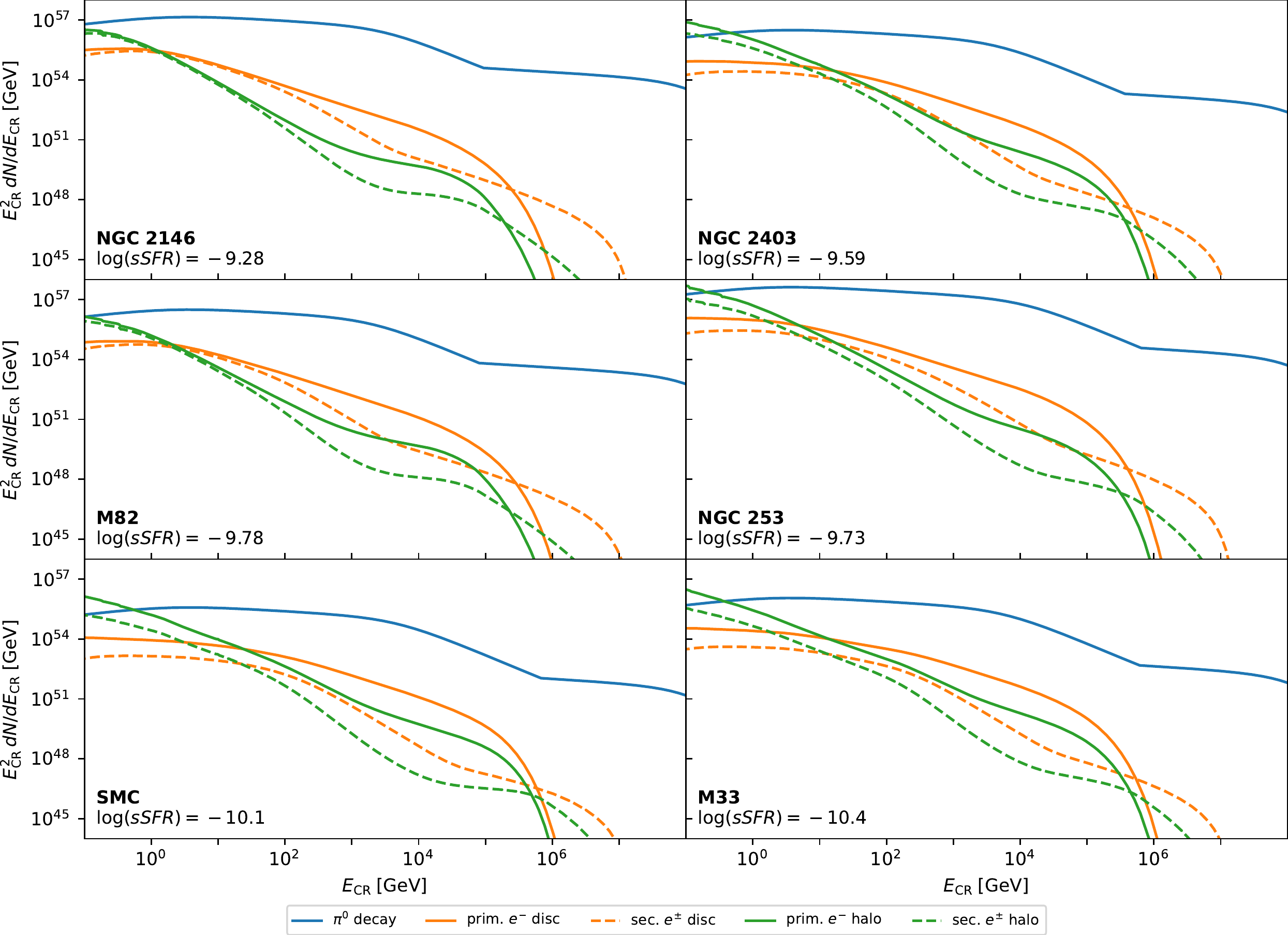}
    \caption{Steady-state CR spectra computed by \textsc{CONGRuENTS} for galaxies across a range of sSFR. Blue lines show the CR proton spectrum in the disc, solid and dashed orange lines show primary and secondary electrons in the disc, and solid and dashed green lines show primary and secondary electrons in the halo.
    }
    \label{fig:CRspectra}
\end{figure*}

\begin{figure*}
	\includegraphics[width=\textwidth]{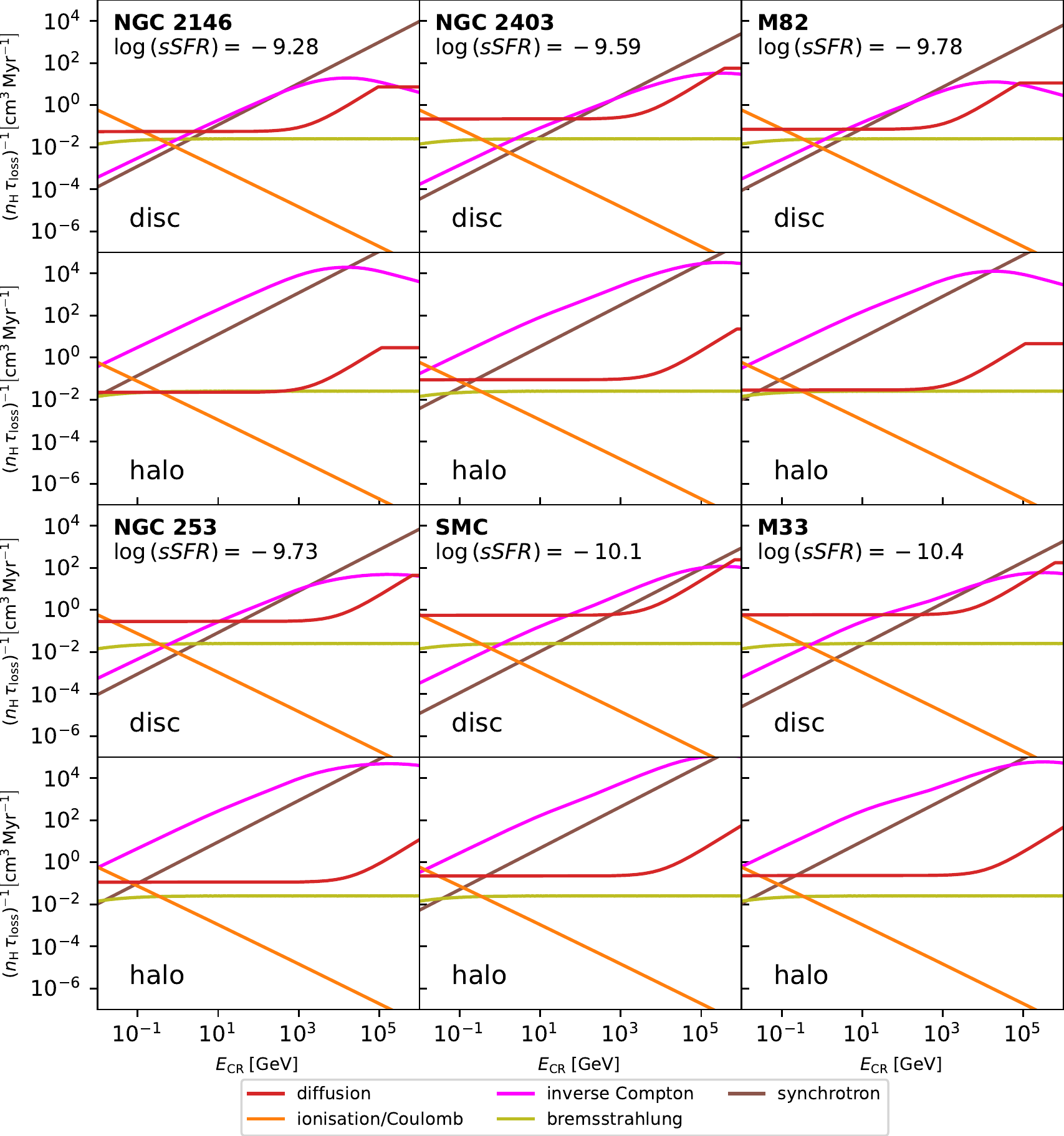}
    \caption{Density-normalised inverse cosmic ray electron loss times for the disc and halo, $\left(n_{\rm H} \, \tau_{\rm loss}\right)^{-1}$. We plot the inverse loss time so that the most important processes appear at the top of each panel, and we normalise by midplane density so that we can compare galaxies with widely different densities on the same scale. Colours indicate the loss process; note that ``ionisation'' is shorthand for ionisation losses in the disc, and Coulomb losses in the halo. The sharp transition to a constant loss time with increasing energy for diffusion for some galaxies is an artefact of the model, and corresponds to the transition where streaming becomes limited by the speed of light. Physically this transition can be expected to be smooth, the details of which have not been modelled.
    }
    \label{fig:tauloss}
\end{figure*}

\begin{figure*}
	\includegraphics[width=\textwidth]{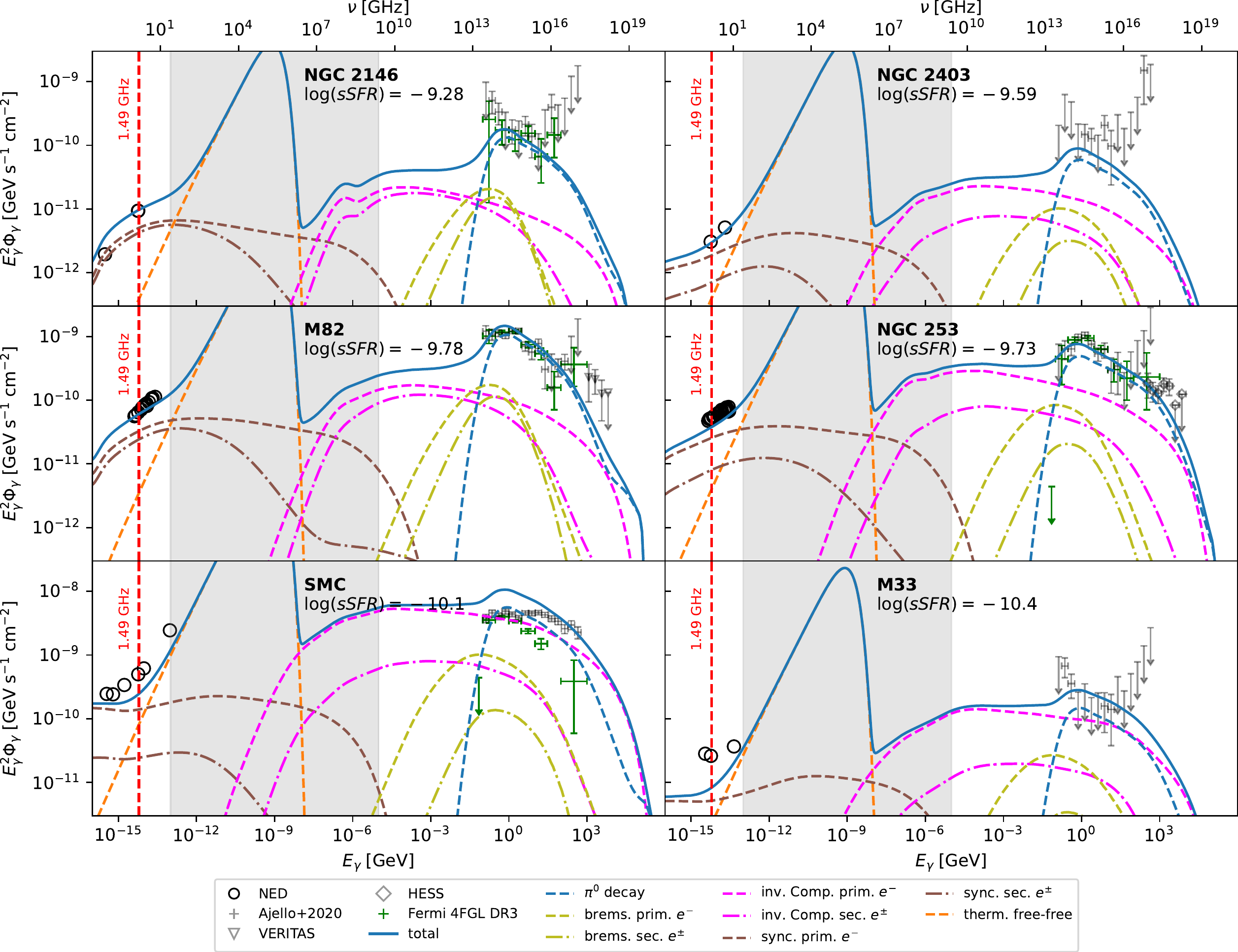}
\caption{Non-thermal galaxy spectra computed using the described model for nearby $\gamma$-ray and radio observed galaxies. The greyed out area is dominated by thermal emission not modelled here. Radio data are taken from the NASA/IPAC Extragalactic Database, $\gamma$-ray data from \citet{Ajello20a} and the Fermi 4FGL DR3 catalogue \citep{2022ApJS..260...53A}, VERITAS data for M82 from \citet{2009Natur.462..770V}, and HESS data for NGC 253 from \citet{2018A&A...617A..73H}.}
    \label{fig:fourgals}
\end{figure*}

We find that our models agree very well with the observations for all the galaxies in our sample except Arp 220, the Large Magellanic Cloud and M31; we defer a discussion of these galaxies to \autoref{ssec:problem_galaxies}. Here we first focus on the other six galaxies in the sample. For each of them, we first compute the ISRFs and show the results in \autoref{fig:radfields}. High star formation rate systems display a spectrum that is dominated (in terms of photon number and total energy) by the FIR emission from reprocessed starlight. In contrast, quiescent systems are dominated by the CMB by number of photons and by the sum of the high energy (3000 K BB and above) components by total energy.

We then proceed to calculate the steady state CR spectra within the disc and the halo, and plot the results in \autoref{fig:CRspectra}. We also show the loss times for CR electrons in each galaxy's disc and halo in \autoref{fig:tauloss}, where we define the loss time as $\tau_{\rm (ion,Coul,sync)} = E/\dot{E}$ for ionisation, Coulomb, and synchrotron losses (which are continuous), $\tau_{\rm (bs,iC)} = 1/\int (d\Gamma/dE_{\rm f}) \, dE_{\rm f}$ for bremsstrahlung and inverse Compton losses (which are catastrophic), and $\tau_{\rm diff} = h_{\rm g}^2/D$ for diffusion out of the disc. Note that, although we do not show protons in the figure to avoid clutter, the diffusive loss times for protons are nearly identical to those of electrons; they differ from the electron ones only at energies $\lesssim 1$ GeV, where electrons and protons of the same kinetic energy have different rigidities. 

The results for protons are essentially the same as those obtained by \citet{2020MNRAS.493.2817K} and \citet{2021Natur.597..341R}, since the underlying model is the same. The degree of calorimetry varies from galaxy to galaxy (\autoref{fig:Sigmagasfcal}) and as a function of proton energy. The curvature and inflections visible in the proton spectra visible at energies in the $\sim \mbox{TeV} - \mbox{PeV}$ range, depending on the galaxy, are a result of the energy dependence of the diffusive loss time. At low energies the loss times are nearly energy-independent, since all protons stream at speeds very close to the ion Alfv\'en speed, but at higher energies the loss times drop as the energy densities decrease and the streaming speeds increase, causing curvature in the spectral shape. At sufficiently high energies the streaming speed saturates at $c$, causing the high-energy inflections visible in the spectra.
Turning to electrons, we find that the discs of
the starbursts with large sSFR are mostly calorimetric (i.e., $\tau_{\rm diff}$ is not the shortest timescale, so electrons lose most of their energy before escaping the disc) over the majority of the electron energy range, but that the energy range over which diffusion is the most rapid process, and thus there is significant escape into the halo, expands for galaxies closer to the star-forming main sequence (c.f. \autoref{fig:SFMS}).
For all systems, CR electron losses at low energies are dominantly by ionisation and Coulomb interactions in the disc and halo, respectively, and scale approximately as $\tau_{\rm ion} \propto R_{\rm e}^{2} \dot{M}_{\ast}^{-1.6} M_{\ast} E_{\rm e}$ -- this scaling follows simply from the functional dependence of $n_{\rm H}$ on $R_{\rm e}$, $\dot{M}_*$, and $M_*$ through the scaling relations in \autoref{ssec:disc}, and the very weak dependence of the ionisation loss rate on CR electron energy (c.f.~\autoref{eq:edot_ion}). At intermediate energies, $\sim 0.1$ GeV to $\sim 10$ GeV, bremsstrahlung losses, for which our adopted empirical scaling relations imply a loss time that scales as $\tau_{\rm bs} \propto R_{\rm e}^{2} \dot{M}_{\ast}^{-1.6} M_{\ast}$, constitute a significant but sub-dominant loss channel in essentially all galaxies.

At still higher energies, $\gtrsim 10$ GeV, synchrotron losses compete with inverse Compton losses.
Again making use of the scaling relations derived in \autoref{ssec:disc} and \autoref{ssec:halo}, the loss times for these two scale, respectively, as $\tau_{\rm sync} \propto R_{\rm e}^{2} \dot{M}_{\ast}^{-2} M_{\ast} E_{\rm e}^{-1}$ and $\tau_{\rm iC} \propto R_{\rm e}^{3} \dot{M}_{\ast}^{-1.7} M_{\ast}^{0.5} E_{\rm e}^{-1}$ (in the Thomson regime) which together imply a weak increase in the ratio of synchrotron to inverse Compton losses ($\propto R_{\rm e} \dot{M}_{\ast}^{0.3} M_{\ast}^{-0.5}$) as the star-formation rate increases and the stellar mass decreases. Synchrotron losses become clearly dominant
once inverse Compton is suppressed in the Klein-Nishina regime. In \autoref{fig:tauloss}, this effect manifests as the curvature in $\tau_{\rm iC}$ at large energies. We note that this effect would not be captured in earlier attempts to model galactic non-thermal spectra that do not properly include the frequency-dependence of the radiation field and its variation with galactic properties -- note for example that the energy at which Klein-Nishina effects become apparent in \autoref{fig:tauloss} is much lower in low sSFR galaxies such as SMC or M33 than high sSFR galaxies such as NGC 253 or NGC 2164, simply because the photon field in the starburst systems is dominated by lower-energy IR photons while that in the weakly star-forming galaxies is dominated by the harder starlight radiation field. Despite this, the Klein-Nishina effect is most important for starburst systems, where the magnetic field amplitudes are larger (and thus synchrotron losses are more important in general), and where the high calorimetry fraction for protons provides a significant population of very high-energy secondary electrons that, thanks to the Klein-Nishina effect, deposit almost all of their energy into the synchrotron rather than inverse Compton loss channel.

Finally, diffusive escape of CR electrons into the halo is significant if the galactic star formation rate is low enough, with the highest escape rates occurring at intermediate energies where the loss times for synchrotron and inverse Compton (dominant at high energy) and ionisation (dominant at low energy) are not too short. As a result, the halo population of cosmic ray electrons is largely truncated to below TeV energies (\autoref{fig:CRspectra}) with the exception of the very lowest luminosity systems where diffusive escape from the disc at high energies is still possible due to the suppression of all other loss channels. Once in the halo, electrons become fully calorimetric in all galaxies, depositing their remaining energy into the synchrotron or inverse Compton channels.
We plot the non-thermal spectra that \textsc{CONGRuENTS} predicts for our sample galaxies in \autoref{fig:fourgals}. The figures show that the models provide a good fit to the observed data at both radio and at $\gamma$-ray energies; the models generally match the data at the tens of percent level.\footnote{We discount here the discrepancies for the $\gamma$-ray spectrum of the SMC, where the observations themselves clearly suffer significant systematic uncertainty, as highlighted by the very large divergence between the \citet{Ajello20a} and Fermi 4FGL DR3 \citep{2022ApJS..260...53A} spectra. These differ from each other by significantly more than our model differs from either one.}
While highly-tailored models can achieve better agreement, here we remind readers that we achieve this level of agreement without using \textit{any} input data other than these galaxies' star formation rates, stellar masses, and size, and without having \textit{any} tuneable free parameters that we can vary from one galaxy to another. A single model fits all the galaxies shown in \autoref{fig:fourgals} simultaneously.

\subsection{Problem galaxies}
\label{ssec:problem_galaxies}

\begin{figure*}
	\includegraphics[width=\textwidth]{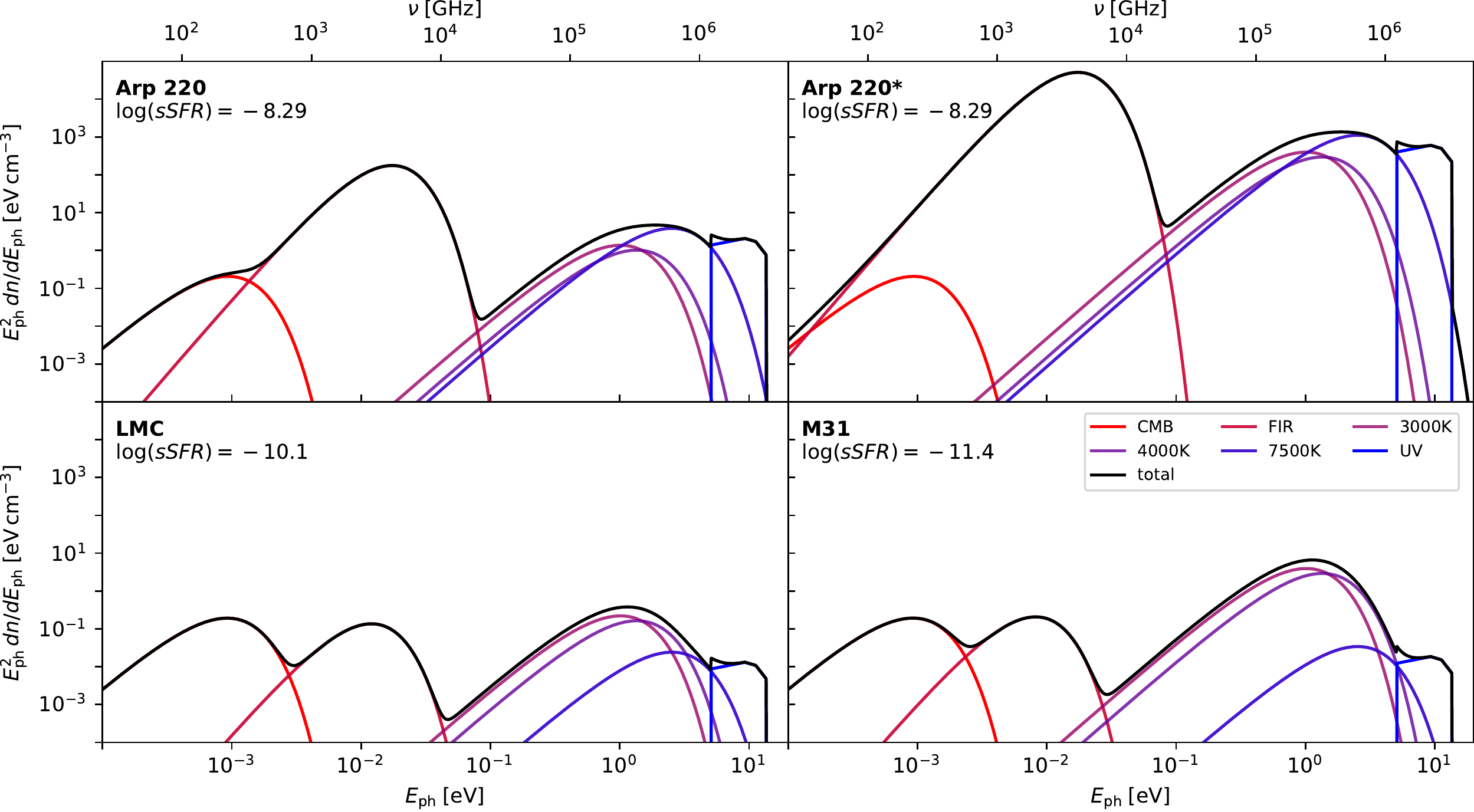}
    \caption{Same as \autoref{fig:radfields}, but for the three galaxies (Arp 220, LMC, M31) that require corrections. The panel for Arp 220* shows the shape of the spectrum after correcting the radius, while the left panel shows the results before correction -- see \autoref{sssec:arp220} for details.}
    \label{fig:radfieldsnsg}
\end{figure*}

\begin{figure*}
	\includegraphics[width=\textwidth]{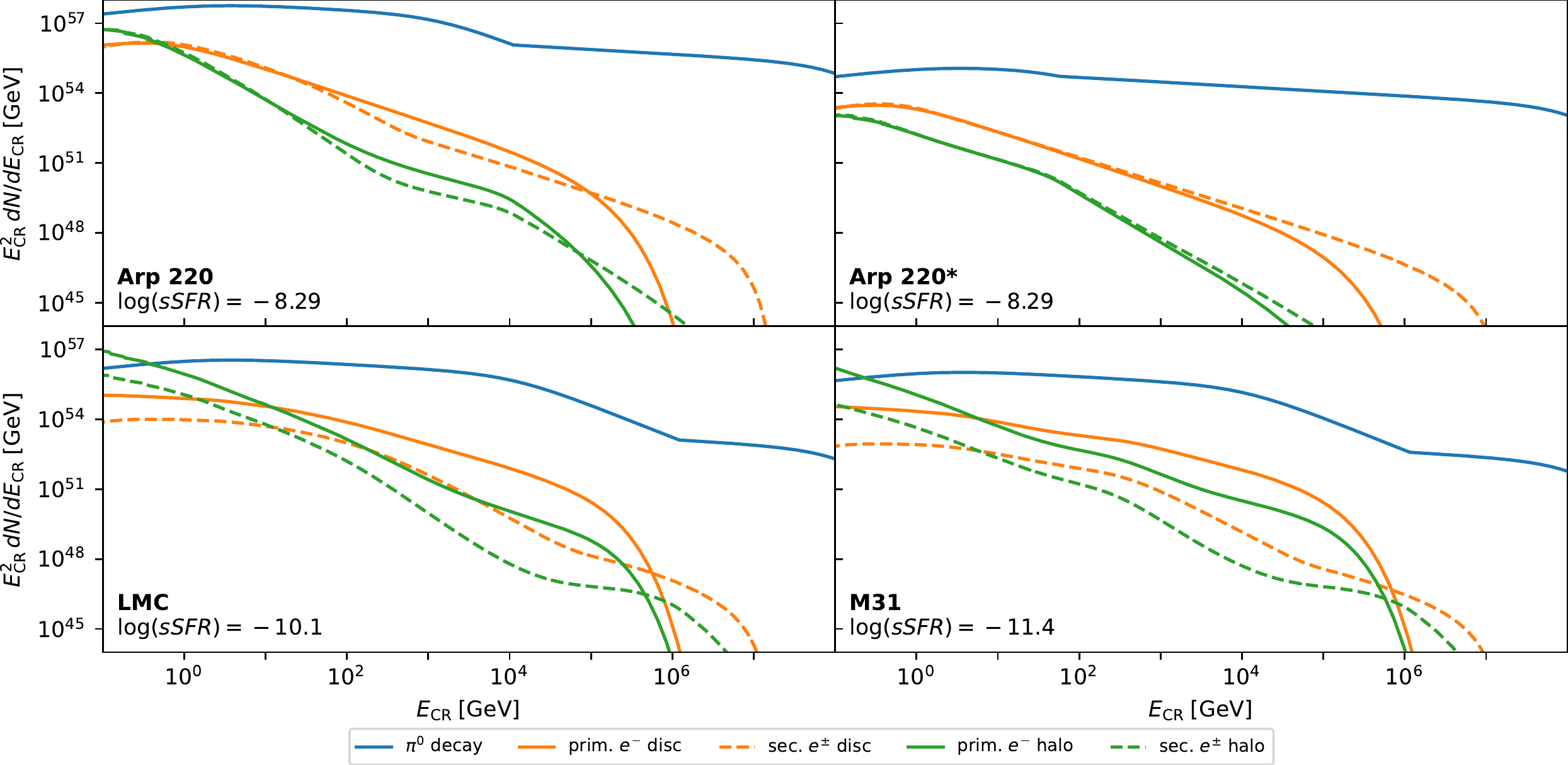}
    \caption{Same as \autoref{fig:CRspectra}, but for the three galaxies (Arp 220, LMC, M31) that require corrections. The panel for Arp 220* shows the shape of the spectra after correcting the radius, while the left panel shows the results before correction, see \autoref{ssec:problem_galaxies}.}
    \label{fig:CRspectransg}
\end{figure*}

\begin{figure*}
	\includegraphics[width=0.7\textwidth]{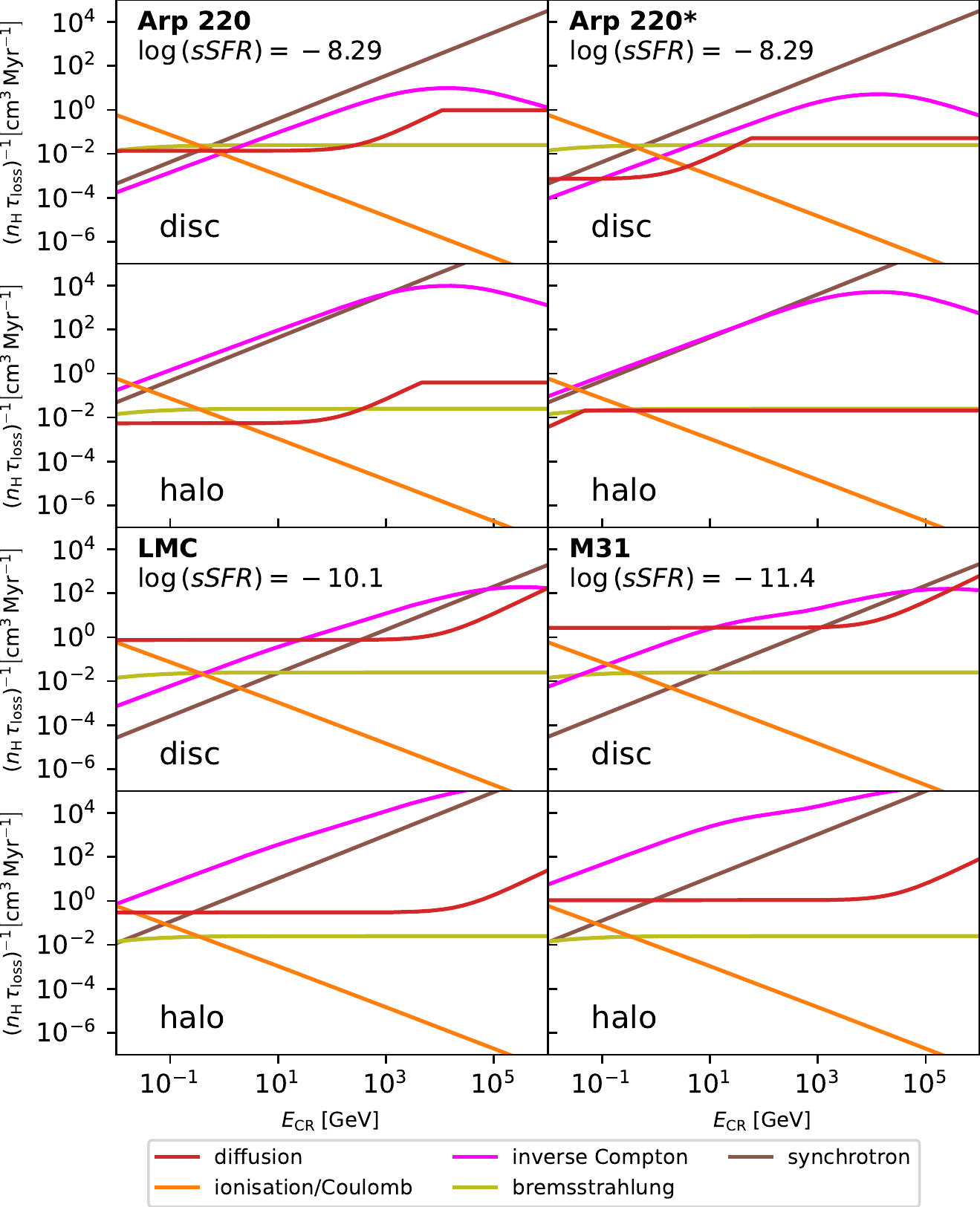}
    \caption{Same as \autoref{fig:tauloss}, but for the three galaxies (Arp 220, LMC, M31) that require corrections. The top left and right panels show the loss times for Arp 220 before and after applying the corrections detailed in \autoref{ssec:problem_galaxies}, respectively.}
    \label{fig:taulossnsg}
\end{figure*}

\begin{figure*}
	\includegraphics[width=\textwidth]{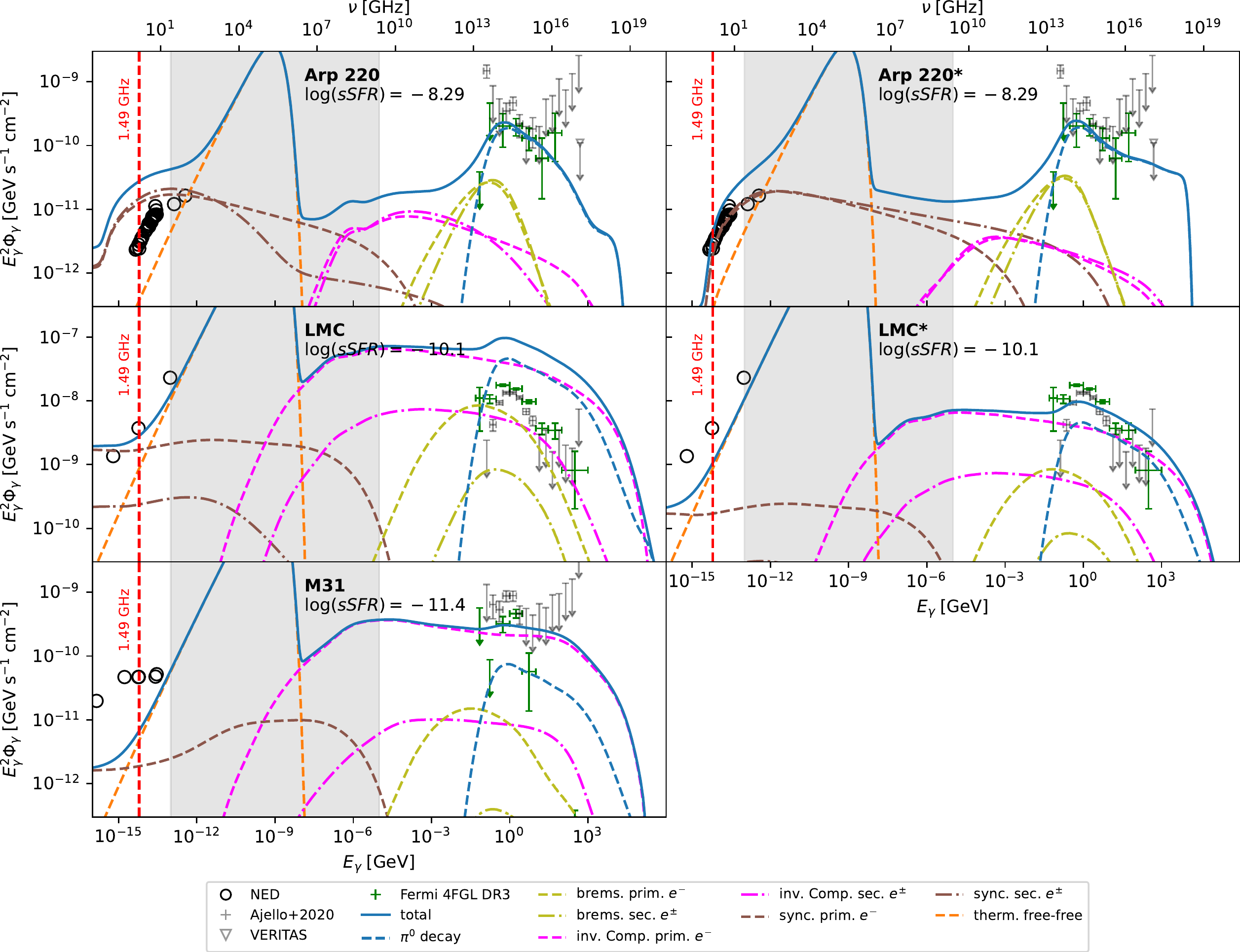}
    \caption{Same as \autoref{fig:fourgals}, but for the three galaxies (Arp 220, LMC, M31) that require corrections. The left column shows our predicted non-thermal spectra before applying the corrections detailed in \autoref{ssec:problem_galaxies}, and the right column shows the results after making corrections. VERITAS data for Arp 220 are taken from \citet{2015ICRC...34..745F}.}
    \label{fig:specsnsg}
\end{figure*}

We next discuss the three cases -- Arp 220, the LMC and M31 -- where naive application of \textsc{CONGRuENTS} yields results that do not match the observed spectrum well. In these cases we are able to identify a plausible physical origin for the discrepancy, and correct for them for Arp 220 and the LMC; we show the ISRFs, steady-state CR spectra spectra, loss times, and non-thermal emission we infer for these galaxies before and after making these corrections in \autoref{fig:radfieldsnsg}, \autoref{fig:CRspectransg}, \autoref{fig:taulossnsg}, and \autoref{fig:specsnsg}, respectively.

\subsubsection{Arp 220}
\label{sssec:arp220}

Naive application of \textsc{CONGRuENTS} to Arp 220 yields a predicted spectrum that is a factor of of $\sim 5-10$ too bright in the radio, as we illustrate in the left column in \autoref{fig:specsnsg}. The reason the model fails is simple: star formation in Arp 220 is concentrated in a heavily dust-enshrouded nuclear starburst. By default \textsc{CONGRuENTS} employs the  optical half-light radius in determining characteristic values for the various local ISM parameters, but for Arp 220 the optical radius is more reflective of the distribution of the old stellar population than the size of the nuclear starburst, which is essentially invisible at optical wavelengths due to its extreme extinction. We can improve the fit dramatically simply by replacing $R_{\rm e}$ in our model with the approximate size of the starburst region as determined from radio observations, $R_{\rm radio} = 200$ pc; while more detailed observations (see e.g. \citealt{2015ApJ...799...10B, 2017ApJ...836...66S}) have resolved the two nuclei at the centre of Arp 220, we adopt our value to approximately enclose the entire starburst region which has previously been measured by \citet{1991ApJ...366L...5S}. Further discussion of the internal structure of Arp 220 and its consequences for the observed emission can be found in \citet{2004ApJ...617..966T, 2013ApJ...762...29L, 2017MNRAS.469L..89Y, 2019MNRAS.484.3665Y}.

We show the revised prediction we generate from this change in the right column in \autoref{fig:specsnsg}, which agrees much better with the observations. The primary difference is a significantly greater free-free opacity that extends to higher energies and suppresses the synchrotron emission. We also show the changes in the shape of the ISRF, predicted CR spectra, and calculated loss times in \autoref{fig:radfieldsnsg}, \autoref{fig:CRspectransg}, and \autoref{fig:taulossnsg}, respectively. These figures show that, as one might expect, reducing the radius has the effect of making the ISRF much more intense and also increasing the density and magnetic field strength, dramatically reducing the loss times and therefore suppressing the steady-state CR spectrum considerably. However, these effects are less important for the final emitted non-thermal spectrum than the change in free-free opacity, simply because Arp 220 is so dense that even using an incorrectly large radius yields essentially full calorimetry for both protons and electrons.

This result points to one potential failure mode of our model: it can make poor predictions for galaxies where the effective radius is not representative of the true radial extent over which star formation occurs and thus CRs are injected. Such galaxies likely require an alternative estimate of the radius. While this is a potential concern for many nuclear starbursts, the fact that our model does well with NGC 2146, M82, and NGC 253 suggests that the problem is moderate, and that there should be no major problems in applying  our model to optical data for main sequence galaxies at $z\gtrsim 1$, which tend to have star formation rates and dust extinctions more similar to those of NGC 2146, M82, and NGC 253 than to Arp 220. However, this result does also suggest that, for galaxies where wavelengths other than optical provide more reliable estimates of the size of the star-forming area, and those multiwavelength data are available, it is preferable to use them for safety.

\subsubsection{The Large Magellanic Cloud}

Naively applying \textsc{CONGRuENTS} to the Large Magellanic Cloud over-predicts the observed $\gamma$-ray emission by an order of magnitude, as shown in \autoref{fig:specsnsg} (left). This is a problem also seen in sophisticated MHD simulations \citep{2020ApJ...893...29B}, assuming currently observed rates of star-formation. Fundamentally, the problem is that the proton calorimetry fraction that our model derives, consistent with the results of most MHD simulations, is $\sim 10\%$, whereas the observed ratio of $\gamma$-ray luminosity to star formation rate in the LMC suggests a figure closer to $\sim 1\%$ \citep[e.g.,][]{Crocker21a}. Even this may be an overestimate: the observed LMC $\gamma$-ray spectrum has a characteristic bump around 1 GeV that does not look like a classical pion spectrum (for plausible cosmic ray spectral index), but instead looks suspiciously like the spectrum associated with prompt millisecond pulsar emission \citep{2022NatAs...6.1317C}. Consistent with the latter scenario, the dominant $\gamma$-ray component appears to be mostly coming from an extended source, rather than confined to regions of active star-formation \citep{2016A&A...586A..71A}. If the observed $\gamma$-rays are largely driven by millisecond pulsars and not CRs, then the CR contribution and the calorimetry level must be even smaller.

We can get an improved match between our \textsc{CONGRuENTS} model and the observed LMC spectrum if we arbitrarily lower the proton calorimetry fraction to 1\%. We show the result of doing so as the ``corrected'' model in the right panel of \autoref{fig:specsnsg}. One possible physical explanation for why reduced calorimetry might be justified is that \citet{2009AJ....138.1243H} find that the star-forming regions 30 Dor, Constellation III and the Northwest Void, combined, make up approximately 45\% of the current star-formation rate of the entire LMC, and that these complexes are quite young, implying that the LMC's star formation rate has increased substantially in the last $\approx 50$ Myr. Since core collapse SNe occur only $\approx 3-40$ Myr after star formation, with half delayed by $\approx 20$ Myr or more \citep[e.g.,][]{Matzner02a}, the CR population in the LMC might not yet have reached the equilibrium value implied by its present-day star formation rate. More generally, non-equilibrium CR populations may be a significant effect in other low luminosity systems where the overall low rate of star formation implies that stochastic fluctuations in the massive star population are non-negligible \citep{2014MNRAS.444.3275D}. Such systems may spend significant portions of their lives with CR populations that are out of equilibrium with their instantaneous SFRs.

\subsubsection{M31}

Our result for Andromeda significantly under-predicts the observed radio emission from the galaxy, and also misses a bump in $\gamma$-ray emission at $\approx 1$ GeV. There are a number of reasons why this might be the case. Andromeda features a massive, old stellar bulge that is not forming stars; star formation is confined to an annulus at a galactocentric radius of around 10 kpc. As discussed previously, our model estimates the interstellar radiation field from the stellar mass and radius, and in the case of M31, where most star formation is far from the galactic centre, this assumption likely leads us to significantly overestimate the strength of the ISRF in the location where star formation is actually occurring. Conversely, due to significant diffusion from their sources of acceleration, energy losses by CR electrons are not confined to the star-forming ring but instead occur throughout the disc and halo of M31.  Our modelled magnetic field strength of a few $\mu$G for the star-forming region is in good agreement with that measured there, but measurements of the magnetic field strength in the inner disc suggest values a factor of several higher \citep{1998IAUS..184..351H,2014A&A...571A..61G,2019PhRvD.100b3014M}. Both errors would lead us to overestimate inverse Compton losses and underestimate synchrotron ones. Thus M31 may be a case similar to Arp 220, where our model oversimplifies because it characterises the size of the star-forming disc by a single number.

However, a second possibility is that the emission from M31 is not driven by star formation at all (or at least not by recent star formation). M31 is likely home to a significant MSP population \citep{2017ApJ...836..208A, 2018ApJ...862...79E, 2022NatAs...6.1317C}, as the characteristic $\sim$GeV bump in the $\gamma$-ray spectrum would suggest, and its galactic centre features a bright source of radio emission that is not associated with star-formation, e.g. see the radio maps in Figure 2 of \citet{2013A&A...557A.129T}. This would suggest that a significant fraction of the radio emission in M31 is not star-formation related but comes from a different source at M31's galactic centre instead.

\section{Summary and Conclusions}
\label{sec:conclusion}

We introduce Cosmic-ray, Neutrino, Gamma-ray and Radio Non-Thermal Spectra (\textsc{CONGRuENTS}), a predictive model for non-thermal emission from star-forming galaxies that relies only on three easily measurable parameters -- the stellar mass, star-formation rate, an effective radius -- and yields good predictions for the spectra of non-thermal emission produced by cosmic ray electrons and protons. Our calculations are based on a physically-motivated model for CR proton transport, and for CR electrons include a full solution to the kinetic equation that properly accounts for catastrophic loss processes. Our treatment of the galactic environment, particularly the magnetic field and interstellar radiation field, is significantly more realistic than past models, and enables us to properly account for effects like the correlation between magnetic field strength and star formation rate, and the shift from optical-dominated radiation fields in low surface density galaxies to IR-dominated ones in dense galaxies. 

We compare our model to $\gamma$-ray and radio observations for a range of local galaxies that sit below, on, and above the star-forming main sequence. We show that for the majority of these galaxies the \textsc{CONGRuENTS} prediction matches the data to within tens of percent, with no inputs other than the three measured parameters listed above, and no parameters that can be tuned differently from one galaxy to another. 

Of the three galaxies where agreement is poorer, we find that the disagreement for two of them can be resolved by simple tweaks -- in the case of Arp 220, using a radius taken from radio rather than optical measurements (which are unreliable due to extreme extinction) and, in the case of the Large Magellanic Cloud, recognising that the recent uptick in the star formation rate implies a cosmic ray population that has not reached the equilibrium that our model assumes.
For the remaining and most recalcitrant case, M31, the ring-like morphology of the star formation, together with substantial non-thermal emission from its large bulge -- which does not correlate with star formation  -- largely defeats our simple assumptions.
The silver lining here is that M31 presents a case where the failure of our model
can be used to identify a situation where, apparently, agents other than star formation
are contributing substantially to the galaxy's non-thermal emission. 
Millisecond pulsars are a natural candidate here
\cite[cf.,][]{2017ApJ...836..208A, 2018ApJ...862...79E,Sudoh2021,Gautam2022,2022NatAs...6.1317C}, an idea that we intend to investigate further in future work.

More broadly, our ability to reproduce observations across a wide range of environments suggests that \textsc{CONGRuENTS} is a valuable tool for making ``first look'' predictions of galaxies' non-thermal emission, which can be made without the need for an extensive suite of observational inputs or an expensive campaign of simulations or observations.

Our models also provide new insight into the nature of CR electron emission in the galaxies we study. We find that all the systems under consideration are CR electron calorimeters, but the loss processes that dominate depend on the electron energy range. Loss processes in the intermediate energy range ($0.1 \ {\rm GeV} \lesssim E_{\rm e} \lesssim 10 \ {\rm GeV}$) depend sensitively on the ISM environment, and are different in galaxies with different properties, whereas ionisation losses generally dominate at low energy and synchrotron and inverse Compton emission at high energy, in a ratio that also depends on galactic environment. At the highest energies, $\gtrsim 1-10$ TeV depending on the galaxy, synchrotron emission always dominates because inverse Compton losses are suppressed by the Klein-Nishina effect. We also find that, for systems with high star-formation rates and associated high proton calorimetry fractions, secondary cosmic ray electrons can contribute up to half of the total observed emission, particularly at high energies.

In a forthcoming series of papers we intend to apply \textsc{CONGRuENTS} to a large sample of galaxies drawn from optical surveys, in order to understand the origins of empirical relations such as the FIR-radio and FIR-$\gamma$ correlations, predict observable extragalactic neutrino emission, and examine the large-scale structure of the non-thermal sky.

\section*{Acknowledgements}

This research has made use of the NASA/IPAC Extragalactic Database, which is funded by the National Aeronautics and Space Administration and operated by the California Institute of Technology. This work has made use of DustPedia, a collaborative focused research project supported by the European Union under the Seventh Framework Programme (2007-2013) call (proposal no. 606847). The participating institutions are: Cardiff University, UK; National Observatory of Athens, Greece; Ghent University, Belgium; Université Paris Sud, France; National Institute for Astrophysics, Italy and CEA, France. MRK and RMC acknowledge support from the Australian Research Council through its \textit{Discovery Projects} funding scheme, award DP190101258, and the Australian National University through a research scholarship (MAR).

\section*{Data Availability}

The data underlying this article are available in Zenodo, at https://doi.org/10.5281/zenodo.7805110. Where external data were used, these are publicly accessible and are referenced in the text.

\section*{Code Availability}

\textsc{CONGRuENTS}, the code used to compute the results of this research is open source and available at https://doi.org/10.5281/zenodo.7935188.



\bibliographystyle{mnras}
\bibliography{bibliography} 




\begin{appendix}

\section{Alternate diffusion models}
\label{app:vai}

\begin{figure*}
	\includegraphics[width=\textwidth]{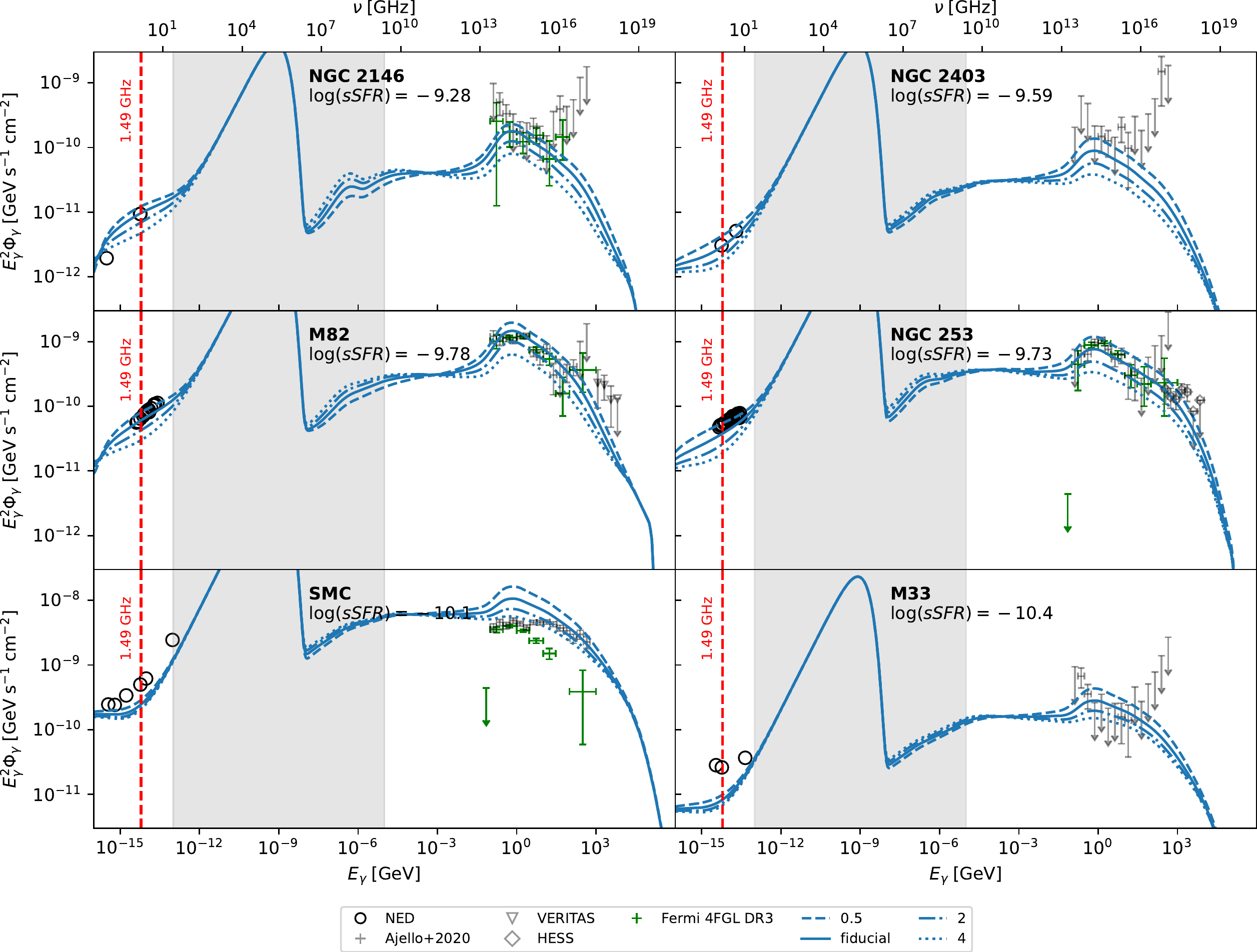}
\caption{Total non-thermal galaxy spectra computed as in \autoref{fig:fourgals} but with $f_{\rm st} = 0.5$ (dashed), 1 (solid; fiducial choice in the main text), 2 (dot-dashed), and 4 (dotted).}
    \label{fig:goodvai}
\end{figure*}

As discussed in the main text, \textsc{CONGRuENTS} allows uses to manually set $f_\mathrm{st}$, the ratio of the streaming speed to the ion Alfv\'en speed, to values other than unity. Here we explore the effects of doing so, using example values $f_\mathrm{st}=0.5$, 2, and 4. In \autoref{fig:goodvai} we show the total spectrum we predict for the set of galaxies shown in \autoref{fig:fourgals} for these alternative values of $f_\mathrm{st}$. From the results we can see that while the change in the radio emission is slight, the $\gamma$-ray spectra either overproduce the observed emission for $f_{\rm st} = 0.5$ in some galaxies, and under-predict in some cases where $f_{\rm st} = 2$ or 4. We can understand this as follows: by increasing $V_{\rm st}$, we reduce the residence time of CRs within the disc, hence we observe significantly lower emission from both CR protons and leptons. For the leptons, however, this is partially compensated: increased escape into the halo for CR leptons leads to more halo synchrotron emission. However, because the ratio of magnetic to radiation energy density is lower in the halo than in the disc, the compensation is only partial, and more lepton energy goes into inverse Compton losses. This leads to the slight uptick of the inverse Compton spectrum at lower energies that we observe at higher $f_\mathrm{st}$.

\section{Numerical method for the electron kinetic equation}
\label{app:algorithm}

Here we describe our method for solving \autoref{eq:electron_spectrum}. We begin by making a change of variable from $E$ to $\ln E$, since, given the wide range of energies, it is more natural to solve for the spectrum per unit log energy rather than per unit energy, and we transform the problem to be conservative in energy. We therefore define $\alpha_{\rm e} = dN_{\rm e} / d\ln E = E q_{\rm e}$, and multiply \autoref{eq:electron_spectrum} by $E^2$, one factor of $E$ for the change of variable and one for the energy conserving scheme. After applying the chain rule to the first term, this yields
\begin{eqnarray}
    0 & = & - \frac{\partial}{\partial \ln E}\left(\dot{E}\alpha_{\rm e}\right) + \dot{E}\alpha_{\rm e} - \frac{D E}{h_{\rm g}^2} \alpha_{\rm e} + Q E^2
    \nonumber
    \\
    & & \qquad {}
    - E \alpha_{\rm e} \int_{\ln m_{\rm e} c^2}^{\ln E} \frac{d\Gamma}{d\ln k}\!(E)\, d\ln k
    \nonumber \\
    & & \qquad {}
    + E \int_{\ln E}^\infty \alpha_{\rm e}(k) \frac{d\Gamma}{d\ln E}\!(k)\, d\ln k.
    \label{eq:alpha_eqn}
\end{eqnarray}

The next step is to discretise using a conservative finite volume method. Consider a set of $N$ energy bins, where $E_{i-1/2}$ and $E_{i+1/2}$ represent the lower and upper limits of energy bin $i$ for $i = 1\ldots N$, and the bins are uniformly spaced in logarithm, i.e., $\ln( E_{i+1/2}/E_{i-1/2}) = \Delta \ln E$ is the same for all $i$. For all computations shown in this work we use $N = 500$. We now average both sides of \autoref{eq:alpha_eqn} over $\ln E$ from $\ln E_{i-1/2}$ to $\ln E_{i+1/2}$, giving
\begin{eqnarray}
    \label{eq:alpha_discrete_1}
    0 & = & -\frac{1}{\Delta\ln E}\left[\left(\dot{E}\alpha_{\rm e}\right)_{i+1/2} - \left(\dot{E}\alpha_{\rm e}\right)_{i-1/2}\right]
    \nonumber
    \\
    & & {} + \left\langle \dot{E}\alpha_{\rm e} \right\rangle_i
    - \frac{\left\langle D E \alpha_{\rm e}\right\rangle_i}{h_{\rm g}^2}
    + \left\langle Q E^2\right\rangle_i
    \nonumber
    \\
    & &
    {} - \left\langle E \alpha_{\rm e} \int_{\ln m_{\rm e} c^2}^{\ln E} \frac{d\Gamma}{d\ln k}\!(E) \,d\ln k \right\rangle_i
    \nonumber \\
    & &
    {} + \left\langle E \int_{\ln E}^\infty \alpha_{\rm e}(k) \frac{d\Gamma}{d\ln E}\!(k)\, d\ln k\right\rangle_i,
\end{eqnarray}
where for any quantity $X$ we define
\begin{equation}
    \langle X\rangle_i \equiv \frac{1}{\Delta\ln E}\int_{\ln E_{i-1/2}}^{\ln E_{i+1/2}} X \, d\ln E,
\end{equation}
and for the first two terms in \autoref{eq:alpha_discrete_1} the subscripts $i \pm 1/2$ indicate that the quantity in parentheses is to be evaluated at energy $E_{i\pm 1/2}$.

To produce a second-order accurate scheme, we approximate $\alpha_{\rm e}$ within each energy bin by a piecewise-linear function, whereby we approximate $\alpha_{\rm e}(\ln E)$ for $\ln E_{i-1/2} < \ln E < \ln E_{i+1/2}$ by
\begin{equation}
    \alpha_e = \alpha_i + m_i \left(\ln E - \ln E_{i}\right),
\end{equation}
where for brevity we define $\alpha_i \equiv \langle \alpha\rangle_i$ as the mean particle number in each bin, $\ln E_i = (\ln E_{i-1/2} + \ln E_{i+1/2})/2$ as the bin central energy, and
\begin{equation}
    m_i =
    \left\{
    \begin{array}{ll}
    (\alpha_2-\alpha_1)/\Delta\ln E, & i = 1 \\
    (\alpha_{i+1}-\alpha_{i-1})/2\Delta\ln E, & i = 2\ldots N-1 \\
    (\alpha_{N}-\alpha_{N-1})/\Delta\ln E, & i = N
    \end{array}
    \right.
    \label{eq:slope}
\end{equation}
as the slope within each bin. Note that this approximation is particle number conserving within each bin, since $\langle m_i (\ln E - \ln E_i)\rangle_i = 0$ by construction; also note that we can produce a simpler, first-order accurate scheme simply by setting $m_i = 0$ for all $i$.

Using this linear approximation, we can evaluate all the terms that appear in \autoref{eq:alpha_discrete_1}. Starting with the first term in square brackets, we note that our linear approximation does not guarantee continuity at bin edges, but since $\dot{E}$ is always negative, the upwind direction at edge $i+1/2$ is always $i+1$. Thus we discretise this term as
\begin{eqnarray}
   -\lefteqn{\frac{1}{\Delta\ln E}\left[\left(\dot{E}\alpha_{\rm e}\right)_{i+1/2} - \left(\dot{E}\alpha_{\rm e}\right)_{i-1/2}\right] = {}}
   \\
   & &
   -\dot{E}_{i+1/2} \left(\frac{\alpha_{i+1}}{\Delta\ln E} + \frac{m_{i+1}}{2}\right) + \dot{E}_{i-1/2} \left(\frac{\alpha_{i}}{\Delta\ln E} + \frac{m_{i}}{2}\right).
   \nonumber
\end{eqnarray}
The diffusion term discretises trivially to
\begin{equation}
    \frac{\left\langle D E \alpha_{\rm e}\right\rangle_i}{h_{\rm g}^2} = \alpha_i \frac{\langle D E \rangle_i}{h_{\rm g}^2} + m_i \frac{\langle D E \rangle_i'}{h_{\rm g}^2} \Delta\ln E,
\end{equation}
where for any quantity $X$ we define
\begin{equation}
    \langle X\rangle_i' \equiv \left\langle X\left(\frac{\ln E - \ln E_i}{\Delta\ln E}\right)\right\rangle_i.
\end{equation}

To discretise the final two terms in \autoref{eq:alpha_discrete_1}, representing the rate at which catastrophic collisions remove energy from and add energy to each bin, respectively, it is helpful to break up the integrals over $k$ into ranges corresponding to different energy bins. That is, we rewrite these terms as
\begin{eqnarray}
    \label{eq:rate_out}
    \lefteqn{
    \left\langle E \alpha_{\rm e} \int_{\ln m_e c^2}^{\ln E} \frac{d\Gamma}{d\ln k}\!(E)\, d\ln k \right\rangle_i =
    }
    \\
    & &
    \left\langle \left(\int_{\ln m_e c^2}^{\ln E_{1/2}} E \alpha_{\rm e}(E) \frac{d\Gamma}{d\ln k}\!(E) \, d\ln k
    \right.\right.
    \nonumber 
    \\
    & & {}
    +
    \sum_{j=1}^{i-1} \int_{\ln E_{j-1/2}}^{\ln E_{j+1/2}} E \alpha_{\rm e}(E)\frac{d\Gamma}{d\ln k}\!(E)\, d\ln k
    \nonumber
    \\
    & & 
    \left.\left.
    {}
    +
    \int_{\ln E_{i-1/2}}^{\ln E} E \alpha_{\rm e}(E)\frac{d\Gamma}{d\ln k}\!(E)\, d\ln k
    \right)\right\rangle_i
    \nonumber
\end{eqnarray}
and
\begin{eqnarray}
    \label{eq:rate_in}
    \lefteqn{
    \left\langle E \int_{\ln E}^\infty \alpha_{\rm e}(k) \frac{d\Gamma}{d\ln E}\!(k) \, d\ln k
    \right\rangle_i
    = }
    \\
    & &
    \left\langle
    E \int_{\ln E}^{\ln E_{i+1/2}} \alpha_{\rm e}(k) \frac{d\Gamma}{d\ln E}\!(k) \, d\ln k
    \right.
    \nonumber \\
    & &
    \left.
    {}
    +
    E \sum_{j=i+1}^N \int_{\ln E_{j-1/2}}^{\ln E_{j+1/2}}
    \alpha_{\rm e}(k) \frac{d\Gamma}{d\ln E}\!(k)\, d\ln k
    \right\rangle_i.
    \nonumber
\end{eqnarray}
Note that, in writing out the final equation, we assume that $\alpha_{\rm e}$ is negligibly small for energies $k > E_{N+1/2}$, i.e., the number of electrons at energies above the maximum energy in our grid can be approximated as zero.
Substituting our piecewise-linear form into the terms gives
\begin{eqnarray}
    \lefteqn{
    \left\langle -E \alpha_{\rm e}(E)\int_{\ln m_e c^2}^{\ln E}  \frac{d\Gamma}{d\ln k}\!(E)\, d\ln k \right.
    } 
    \\
    \lefteqn{
    {} +
    \left. E\int_{\ln E}^\infty \alpha_{\rm e}(k) \frac{d\Gamma}{d\ln E}\!(k)\, d\ln k 
    \right\rangle_i
    }
    \nonumber
    \\
    & = &
    -\alpha_i \left[\langle\Gamma E\rangle_{i\to 0} + \sum_{j=1}^{i} \langle\Gamma E\rangle_{i\to j}\right]
    + \sum_{j=i+1}^{N} \alpha_j \langle\Gamma E\rangle_{j\to i}
    \nonumber
    \\
    & & {}
    - \Delta\ln E \left[
    m_i \left(\langle\Gamma E\rangle'_{i\to 0}
    + \sum_{j=1}^{i} \langle\Gamma E\rangle'_{i\to j}\right) - 
    \sum_{j=i+1}^N m_j \langle\Gamma E\rangle'_{j\to i}
    \right]
    \nonumber
\end{eqnarray}
where we define
\begin{eqnarray}
    \langle\Gamma E\rangle_{i\to 0} & \equiv & \left\langle E \int_{\ln m_{\rm e} c^2}^{\ln E_{1/2}} \frac{d\Gamma}{d\ln k}\!(E) d\ln k\right\rangle_i \\
    \langle\Gamma E\rangle_{i\to j} & \equiv & \left\langle E \int_{\ln E_{j-1/2}}^{\ln E_{j+1/2}} \frac{d\Gamma}{d\ln k}\!(E) \, d\ln k\right\rangle_i \\
    \langle\Gamma E\rangle_{i\to i} & \equiv &
    \left\langle
    E \left[
    \int_{\ln E_{i-1/2}}^{\ln E} \frac{d\Gamma}{d\ln k}\!(E)\,d\ln k - {}
    \right.
    \right.
    \nonumber \\
    & & 
    \left.\left.
    \int_{\ln E}^{\ln E_{i+1/2}}  \frac{d\Gamma}{d\ln k}\!(k)\,d\ln k
    \right]
    \right\rangle_i
    ,
\end{eqnarray}
and similarly for $\langle\Gamma E\rangle'$. The physical meanings of these terms is is simple: $\langle\Gamma E\rangle_{i\to 0}$ is the rate at which catastrophic collisions move energy from bin $i$ to energies below $E_{-1/2}$, the lowest energy we follow. Similarly, $\langle \Gamma E\rangle_{i \to j}$ for $j \neq i$ and $j\neq 0$ is the rate at which catastrophic transitions move energy from bin $i$ to bin $j$, assuming that $\alpha_{\rm e}$ is constant across bin $i$, while $\langle \Gamma E\rangle_{i \to i}$ is the rate at which catastrophic collisions remove energy from bin $i$ in collisions where the energy lost per encounter is not large enough to cause electrons to move to a lower energy bin. All of these terms are computed under the assumption that $\alpha_{\rm e}$ is constant across the bin, while the terms $\langle\Gamma E\rangle'$ represent a first-order correction to this assumption.

Collecting the various terms, our final discretised equation becomes
\begin{eqnarray}
0 & = &
-\dot{E}_{i+1/2} \left(\frac{\alpha_{i+1}}{\Delta\ln E} + \frac{m_{i+1}}{2}\right)
\\
& &
{} + 
\dot{E}_{i-1/2} \left(\frac{\alpha_{i}}{\Delta\ln E} + \frac{m_{i}}{2}\right)
\nonumber
\\
& &
{} + \alpha_i \langle \dot{E} \rangle_i + m_i \langle \dot{E} \rangle_i' \Delta\ln E
- \alpha_i \frac{\langle D E \rangle_i}{h_{\rm g}^2} - m_i \frac{\langle D E\rangle_i'}{h_{\rm g}^2} \Delta\ln E
\nonumber \\
& &
{}
    + \langle Q E^2 \rangle_i - \alpha_i \left[\langle\Gamma E\rangle_{i\to 0} + \sum_{j=1}^{i} \langle\Gamma E\rangle_{i\to j}\right]
    \nonumber \\
    & & {}
    + \sum_{j=i+1}^{N} \alpha_j \langle\Gamma\rangle_{j\to i}
    \nonumber
    \\
    & & {}
    - \Delta\ln E \left[
    m_i \left(\langle\Gamma E\rangle'_{i\to 0}
    + \sum_{j=1}^{i} \langle\Gamma E\rangle'_{i\to j}\right)
    \right.
    \nonumber \\
    & &
    \left.
    \qquad\qquad
    {}
    - 
    \sum_{j=i+1}^N m_j \langle\Gamma E\rangle'_{j\to i}
    \right].
    \nonumber
\end{eqnarray}
This constitutes a linear system of equations for the $\alpha_i$, and thus the problem can be posed as a matrix equation
\begin{equation}
    \mathbf{M}\boldsymbol{\alpha} = \mathbf{S},
\end{equation}
where $\boldsymbol{\alpha}$ is an $(N+2)$-element vector of $\alpha_j$ values ($N$ real elements plus two ghost elements to enforce the boundary conditions on the slope $m$), $\mathbf{S}$ is a vector of $N+2$ source terms with elements $S_i = -\langle Q E^2 \rangle_i$ for $i=1\ldots N$ and $S_0 = S_{N+1} = 0$, and $\mathbf{M}$ is an $(N+2)\times (N+2)$ matrix where for $i = 1\ldots N$ the entries are
\begin{equation}
    M_{ij} =
    \left\{
    \begin{array}{ll}
    0, & i > j+1 \\\\
    -\frac{\dot{E}_{i-1/2}}{4\Delta\ln E} + \frac{\langle DE\rangle'_i}{2h_{\rm g}^2} - \frac{\langle \dot{E} \rangle_i'}{2} \\
    \quad {}
    + \frac{\langle\Gamma E\rangle'_{i\to 0} + \sum_{k=1}^{i}\langle\Gamma E\rangle'_{i\to k}}{2}
    , & i = j+1 \\\\
    \frac{\dot{E}_{i-1/2}}{\Delta\ln E} + \frac{\dot{E}_{i+1/2}}{4\Delta\ln E} - \frac{\langle DE\rangle_i}{h_{\rm g}^2} + \langle \dot{E} \rangle_i
    \\
    \qquad {}
    - \frac{\langle\Gamma E\rangle'_{i+1\to i}}{2} - \langle\Gamma E\rangle_{i\to 0} 
    \\
    \qquad {}
    - \sum_{k=1}^{i} \langle\Gamma E\rangle_{i\to k}, & i = j \\\\
    -\frac{\dot{E}_{i+1/2}}{\Delta\ln E} + \frac{\dot{E}_{i-1/2}}{4 \Delta\ln E} - \frac{\langle DE\rangle'_i}{2h_{\rm g}^2} + \frac{\langle \dot{E} \rangle_i'}{2}
    \\
    \qquad {}
    + \langle\Gamma E\rangle_{i+1\to i} - \frac{\langle\Gamma E \rangle'_{i\to 0} + \sum_{k=1}^{i}\langle\Gamma E \rangle'_{i\to k}}{2} 
    \\
    \qquad {}
    - \frac{\langle\Gamma E\rangle'_{i+2\to i}}{2}, & i = j-1 \\\\
    -\frac{\dot{E}_{i+1/2}}{4\Delta\ln E} + \langle\Gamma E\rangle_{i+2\to i}
    \\
    \qquad {}
    + \frac{\langle\Gamma E\rangle'_{i+1\to i}}{2} - \frac{\langle\Gamma E\rangle'_{i+3\to i}}{2}, & i = j-2 \\\\
    \langle\Gamma E\rangle_{j\to i} + \frac{\langle\Gamma E\rangle'_{j-1\to i}}{2} - \frac{\langle\Gamma E\rangle'_{j+1\to i}}{2}, & i < j-2
    \end{array}
    \right.
\end{equation}
with the convention that the $\langle\Gamma E\rangle_{i\to j}$ and $\langle\Gamma E\rangle'_{i\to j}$ symbols are zero for $i$ or $j$ outside the range $1\ldots N$. The boundary values, which enforce the correct values of $m_1$ and $m_N$ (c.f.~\autoref{eq:slope}), are $M_{00}=M_{N+1,N+1}=1$, $M_{01}=M_{N+1,N}=-2$, $M_{02}=M_{N+1,N-1}=1$, and $M_{0i}=M_{N+1,j}=0$ for $i>2$ and $j<N-1$. The equivalent matrix for a first-order accurate scheme, with all slopes $m_i$ set to zero, is
\begin{equation}
    M_{ij} =
    \left\{
    \begin{array}{ll}
    0, & i > j \\\\
    \frac{(\dot{E})_{i-1/2}}{\Delta\ln E} - \frac{\langle D\rangle_i}{h_{\rm g}^2} + \langle\Gamma\rangle_{i\to 0} 
    \\
    \quad
    - \sum_{k=1}^{i} \langle\Gamma E\rangle_{i\to k}, & i = j \\\\
    -\frac{(\dot{E})_{i+1/2}}{\Delta\ln E} + \langle\Gamma E\rangle_{i+1\to i}, & i = j-1 \\\\
    \langle\Gamma E\rangle_{j\to i}, & i < j-1
    \end{array}
    \right.
\end{equation}
For a first-order scheme we do not require the boundary conditions, and thus $\boldsymbol{\alpha}$, $\mathbf{Q}$, and $\mathbf{M}$ are $N$- rather than $(N+2)$-elements in size.

The coefficients of the linear system depend on the various terms in angle brackets, which depend on the properties of the galaxy (e.g., number density, radiation field, etc.). However, since this dependence is linear for all the $\langle\Gamma E\rangle$ terms, only the averages $\langle DE\rangle_i$, $\langle DE\rangle'_i$, and $\langle QE^2\rangle_i$ (which depend on the shape of the proton spectrum) need to be evaluated separately for each galaxy. The integrals for all the $\langle\Gamma E\rangle$ terms  can simply be evaluated once for fixed values of number density, magnetic energy density, radiation field, etc., and then rescaled to the properties of an individual galaxy. For each galaxy we do so to fill in the entries in $\mathbf{M}$ and $\mathbf{S}$, then solve for $\boldsymbol{\alpha}$ using a standard LU decomposition method.

\end{appendix}



\bsp	
\label{lastpage}
\end{document}